\documentclass[12pt]{article}
\pdfoutput=1
\usepackage{amsfonts,amsmath,amssymb}
\usepackage[paper=letterpaper,margin=1.0in]{geometry}
\usepackage{graphicx}
\usepackage{cite}
\usepackage{chngcntr}
\usepackage{tocloft}
\usepackage{hyperref}
\usepackage{xcolor}

\usepackage[sort,numbers]{natbib}

\newcommand\be{\begin{equation}}
\newcommand\ee{\end{equation}}
\newcommand\bea{\begin{eqnarray}}
\newcommand\eea{\end{eqnarray}}
\newcommand\ba{\begin{array}}
\newcommand\ea{\end{array}}
\newcommand\ben{\begin{enumerate}}
\newcommand\een{\end{enumerate}}
\newcommand\bi{\begin{itemize}}
\newcommand\ei{\end{itemize}}
\newcommand\bc{\begin{center}}
\newcommand\ec{\end{center}}
\newcommand\bfig{\begin{figure}}
\newcommand\efig{\end{figure}}
\newcommand\bq{\begin{quotation}}
\newcommand\eq{\end{quotation}}
\newcommand\bt{\begin{table}}
\newcommand\et{\end{table}}
\newcommand\btab{\begin{tabular}}
\newcommand\etab{\end{tabular}}
\newcommand\nn{\nonumber}
\newcommand\eref[1]{(\ref{#1})}

\newcommand\tr{\mathrm{Tr}}
\renewcommand\thefootnote{\fnsymbol{footnote}}
\newcommand\comment[1]{}
\newcommand\com{\comment}

\renewcommand\tilde{\widetilde}

\newcommand\sn{\mathrm{sn}}
\newcommand\cn{\mathrm{cn}}
\newcommand\dn{\mathrm{dn}}
\newcommand\am{\mathrm{am}}
\newcommand\sech{\mathrm{sech}}

\begin{document}

\vspace{.5cm}

\begin{center}
{\Huge \textbf{On the Shape of Things}}

\vspace{.2cm}

{\Large \textbf{From holography to elastica}}

\vspace{.6cm}

Piermarco Fonda\footnote{\href{mailto:fonda@lorentz.leidenuniv.nl}{\texttt{fonda@lorentz.leidenuniv.nl}}}$^{1,2}$,
Vishnu Jejjala\footnote{\href{mailto:vishnu@neo.phys.wits.ac.za}{\texttt{vishnu@neo.phys.wits.ac.za}}}$^2$,
\'Alvaro V\'eliz-Osorio\footnote{\href{mailto:aveliz@gmail.com}{\texttt{aveliz@gmail.com}}}$^{2,3,4}$

\vspace{.33cm}

{${}^{1}$ Instituut-Lorentz, Universiteit Leiden,\\ P.O. Box 9506, 2300 RA Leiden, The Netherlands\\}

\vspace*{.2cm}
{${}^{2}$ Mandelstam Institute for Theoretical Physics, School of Physics, NITheP, \& CoE-MaSS,\\
University of the Witwatersrand, WITS 2050, Johannesburg, South Africa\\}

\vspace*{.2cm}
{${}^{3}$ Institute of Physics, Jagiellonian University,\\
Lojasiewicza 11, 30-348 Krakow, Poland\\}

\vspace*{.2cm}
{${}^{4}$ Department of Physics, Queen Mary, University of London,\\
Mile End Road, London E1 4NS, UK}
\end{center}

\vspace{1cm}

\centerline{\textbf{Abstract}}
\bigskip

We explore the question of which shape a manifold is compelled to take when immersed in another one, provided it must be the extremum of some functional. We consider a family of functionals which depend quadratically on the extrinsic curvatures and on projections of the ambient curvatures. These functionals capture a number of physical setups ranging from holography to the study of membranes and elastica. We present a detailed derivation of the equations of motion, known as the shape equations, placing particular emphasis on the issue of gauge freedom in the choice of normal frame. We apply these equations to the particular case of holographic entanglement entropy for higher curvature three dimensional gravity and find new classes of entangling curves. In particular, we discuss the case of New Massive Gravity where we show that non-geodesic entangling curves have always a smaller on-shell value of the entropy functional. Then we apply this formalism to the computation of the entanglement entropy for dual logarithmic CFTs. Nevertheless, the correct value for the entanglement entropy is provided by geodesics. Then, we discuss the importance of these equations in the context of classical elastica and comment on terms that break gauge invariance.

\newpage

\setcounter{tocdepth}{2}
\tableofcontents

\setcounter{footnote}{0}
\renewcommand\thefootnote{\arabic{footnote}}

\section{Introduction}\label{sec:intro}

Constrained optimization problems are a persistent leitmotif in the history of mathematics and physics.
The calculus of variations, which yields classical solutions to minimization problems with prescribed boundary conditions, supplies the language for characterizing equilibrium configurations in diverse physical settings.
A class of problems of particular interest in this context comprises the behavior of gravitational systems.
More than a century ago, Einstein and Hilbert deduced that an action constituted out of purely geometric quantities describes how spacetime curves in response to energy and matter.
The equations of motion obtained from variation of the action are the Einstein equations of general relativity.
If we incorporate higher order, though still purely geometric terms into the action, the equations are suitably modified.
This supplies a theoretical basis for organizing the low energy effective action of gravity as an $\alpha'$ expansion.
The philosophy extends to environments in which the energy functional of a system is written in terms of geometric invariants, for example in determining the shapes of elastic membranes.
The goal of this paper is to formulate solutions to constrained optimization problems couched in terms of geometric actions within a unified framework.

We consider immersions of a lower dimensional manifold in a higher dimensional one.
We study the shape that the immersed submanifold takes if we demand that it extremizes a certain effective action.
This effective action is constructed out of intrinsic, ambient, and extrinsic curvatures order by order in a derivative expansion.
The most familiar case of extrema of this kind of functionals are minimal submanifolds, of which geodesics and minimal surfaces are the lowest dimensional instances.
These shapes are ubiquitous in nature, \emph{e.g.}, the latter are physically realized by soap bubbles in open frames.
There is a rich literature on this theme in mathematics (see, for example,~\cite{CM} and references therein).
In this work, the functionals discussed are more complicated than area functionals and support other classes of extrema, such as Willmore submanifolds~ \cite{Willmore1, Willmore2,Marques2013}.
In order to find the equations satisfied by extrema, referred to as \emph{shape equations}, we must perform a careful variational analysis of the effective action.
Many of the tools and results leading to these equations can be found in the literature with varying degrees of generality and using diverse approaches.
(See the references in Section~\ref{shapeeq}.)
Here, we provide our derivation of the equations for rather general setups.
Perhaps the most important thing to keep in mind in deriving the shape equations 
is to be meticulous about how the geometry of the submanifold looks from an intrinsic and from an extrinsic viewpoint.
This perspective will lead to a number of interesting insights such as the existence of a freedom in the choice of normal directions and its consequences. 

Within the context of the gauge/gravity correspondence~\cite{Maldacena1998,Gubser:1998bc,Witten:1998qj}, the Ryu--Takayanagi prescription~\cite{Ryu2006} states that the problem of computing the entanglement entropy of a region in the boundary conformal field theory (CFT) can be reformulated as a question regarding minimal surfaces in anti-de Sitter space (AdS).
Furthermore, if the gravity action receives corrections from a derivative expansion, we can still calculate this quantity using more general functionals of the class discussed in~\cite{Bhattacharyya2014,Camps2014,Dong2014}.
As a matter of fact, it is known that for four derivative gravity, the entanglement entropy can be obtained by evaluating the relevant functional on one of its extrema~\cite{Dong2014}.
However, the issue of which of the possible extrema provides the right answer is yet to be resolved.
For field theories with four derivative gravity duals this functional falls within the class of effective actions we consider, and thus, the shape equation formalism can be applied directly in this context.
One simply needs to consider an asymptotically AdS (AAdS) ambient manifold, tune the coefficients in the effective action properly, and choose appropriate boundary conditions.
Having a detailed knowledge of the shape equations and its space of solutions might be of use in elucidating how to systematically choose the extremum that yields the right value for entanglement entropy, among other things.

Indeed, we shall see that for four derivative gravity in AdS$_3$  finding all the possible extrema analytically is feasible.
In fact, this is just an example of the problem of finding extremal curves in maximally symmetric spaces treated in~\cite{Langer1984} and discussed in detail in this work.
Then, for concreteness one can consider a particular theory of gravity, such as New Massive Gravity~\cite{Bergshoeff:2009hq}.
In this theory, we find by evaluating the functional on all the relevant extrema that the one on which it takes the largest value, the geodesic, provides the correct value for the entanglement entropy. We invite the reader to consider the elegance and effectiveness of this approach. The standard strategy when dealing with these kind of problems has been to directly derive the equations of motion for the extrema without relying on their geometric structure. It might be helpful to compare the results in the present work with references \cite{Erdmenger2014} (see discussions around Fig.~4 and (B.3)), \cite{Hosseini2015} (see~(6.5)) and \cite{Basanisi:2016hsh} (see the discussion around Fig.~1 and~(A.5)), which are representative of the state of the art. The equations resulting from this method are rather convoluted and finding analytic solutions seems extremely difficult. Thus, one was compelled to rely either on numerical methods or trial and error. In contrast, using the geometrical tools discussed in Sec.~3 one is able to find analytically all the possible extrema for the entanglement entropy functional. This is one of the main results of this present paper.

One of the main advantages of taking a geometric approach is that it can be applied in a wide variety of systems.
Presumably, the first framework that comes to mind when considering applications is the dynamics of curves and surfaces immersed in $\mathbb{R}^3$; after all, these geometries are a part of our everyday lives.
Energy functionals, closely related to the effective actions we consider, emerge in interesting problems of elasticity.
We would like to mention two cases, one for surfaces and the other for curves.
The former is the Canham--Helfrich energy,~\eqref{CanhamHelfrich1}, which can be used to model the elastic properties of a lipid bilayer membrane~\cite{citeulike:7528087,Helfrich1973}.
Interestingly, the shape equations corresponding to this energy were used to predict the existence of a lipid torus for which the ratio between the radii is $\sqrt {2}$~\cite{PhysRevA.41.4517}.
Indeed, this prediction was experimentally verified in~\cite{PhysRevA.43.4525}.
The other example we would like to mention is the Sadowsky--W\"underlich energy,~\eqref{Sad Wun}.
This functional estimates the free energy of a thin elastic ribbon in terms of a curve via dimensional reduction to its centerline.
This model can be used to elucidate certain properties of long polymers~\cite{MASY:MASY200450226}.
Above, we were cautious and said that these functionals are closely related to the ones we study.
There is a crucial difference, the energy functionals~\eqref{CanhamHelfrich1} and~\eqref{Sad Wun} allow for the presence of terms that violate gauge invariance.
From the viewpoint of geometric effective actions, adding such terms needs to be justified on physical grounds.
We believe that this is an important point, and we hope that the developments presented here help to streamline the reasoning.

The organization of the paper is as follows.
In Section~\ref{sec:two}, we introduce the general geometric setup, then we discuss the subject of gauge freedom and normal frames; afterwards, we explain how to obtain the effective action and display the shape equations characterizing their extrema.
In Section~\ref{sec:three}, we apply the shape equation formalism to immersions into a maximally symmetric ambient space, paying particular attention to curves immersed in surfaces.
In Section~\ref{sec:four}, we apply these results to study questions regarding holographic entanglement entropy.
We make general observations regarding the choice of entangling curves and discuss holographic entanglement entropy for logarithmic CFTs.
Section~\ref{sec:five} contains remarks concerning gauge freedom and functionals used to describe elastic curves and surfaces in $\mathbb{R}^3$.
Finally, Section~\ref{sec:six} contains a detailed summary of this work and potential directions for further investigation.
Most of the technical details have been placed in the appendices.
In \ref{sec:app geometry}, we develop the geometric technology needed to derive the equations of motion.
Then, \ref{eom} contains the derivation of the shape equations using the tools developed in the previous appendix.
\ref{sec:trK} explains how to invert the extrinsic curvature in maximally symmetric spaces in order to find the shapes of extrema.
Finally, \ref{sec:jac} provides a brief review of the Jacobi elliptic functions.


\subsection{Notation}\label{notation}
For the reader's convenience, we collect the notation used in this paper.
\begin{center}
\begin{tabular}{lll}
  Symbol & Nomenclature & Definition \\
\hline\\
  $\Sigma$ & Immersed space& $\Sigma = \{x^{\mu}(\sigma_i) |  \; i = 1,\dots,p \}$\vspace{1.2mm} \\
  $\mu$, $\nu,\dots$ & Ambient space indices& $\mu=1,\dots,d$ \vspace{1.2mm} \\
  $i$, $j,\dots$ &  Indices tangent to $\Sigma$  & $i=1\dots p$ \vspace{1.2mm} \\
  $A$, $B,\dots$ &   Indices normal to $\Sigma$  &  $A=1\dots d-p$ \vspace{1.2mm} \\
  $t_{i}^\mu$ & Tangent vectors on $\Sigma$  & $t^\mu_i= \partial_i x^\mu $ \vspace{1.2mm} \\
  $h_{ij}$ &  Induced metric on $\Sigma$ & $h_{ij}=g_{\mu\nu} \partial_i x^\mu \partial_j x^\nu$ \vspace{1.2mm} \\
  $\tilde\nabla_i$ & Intrinsic Levi-Civita &   $\tilde\nabla_k h_{ij}=0$ \vspace{1.2mm} \\
  $\tilde\Delta$ & Intrinsic Laplace--Beltrami & $\tilde\Delta= \tilde\nabla^k\tilde\nabla_k$ \vspace{1.2mm} \\
  $\mathcal{R}^l_{\;kji} $ & Intrinsic Riemann tensor & $\mathcal{R}^l_{\;kji} v_l= [\tilde\nabla_i,\tilde\nabla_j] v_k $ \vspace{1.2mm} \\
  $n_{\mu}^A$ & Normal vectors to $\Sigma$ & $n_{A}^\mu t_\mu=0 $ \vspace{1.2mm} \\
  $\eta_{AB}$ & Metric on the normal bundle& $\eta_{AB} =\rm{diag}(-1,\dots,-1,1,\dots,1) $ \vspace{1.2mm} \\
  $K_{ij}^A$ &  Extrinsic curvatures  &$ K_{ij}^A = t_i^\mu t_j^\nu \nabla_\mu n_\nu^A$\vspace{1.3mm} \\
  $T^{AB}_i $ & Extrinsic torsion& $T^{AB}_i =t_i^\mu n^{A\nu} \nabla_\mu n_\nu^B$ \vspace{1.2mm} \\
$\tilde{D}_{i\;\;B}^{\;A} $ &  Gauge covariant derivative &$\tilde{D}_{i\;\;B}^{\;A} V^B_{j\dots}=\tilde{\nabla}_{i} V^A_{j\dots} 
 + 
  T_i^{AB} \eta_{BC} V^C_{j\dots}$\vspace{1.2mm} \\
\end{tabular}
\end{center}


\section{The effective action and shape equations}\label{sec:two}
In this section, we describe how to specify the most general effective action up to quadratic order in the curvatures.
We then write the corresponding equations of motion.

\subsection{Geometric setup}\label{geo setup}
We start by considering an immersion
\bea
f:\ N &\to& M \nn \\
\sigma^i &\mapsto& x^\mu(\sigma^i) ~. \label{eq:imm}
\eea
The manifold $N$ is $p$ dimensional, so that a point $P\in N$ is specified by coordinates $\sigma^i$, $i=1,\ldots,p$.
The map $f$ takes $P$ and sends it to the point $f(P)\in M$.
Thus, if $M$ is $d$ dimensional, we may write coordinates $(x^1,\ldots,x^d)$ for $f(P)$.
We observe that each of the $x^\mu$, $\mu=1,\ldots,d$, are functions of the coordinates on $N$. We define $\Sigma\subset M$ to be the orientable submanifold obtained from taking the images of all of the points $P\in N$ under the map~\eref{eq:imm}:
\be
\Sigma = f(N) \subset M ~.
\ee
When $N$ is diffeomorphic to its image $\Sigma$, then $f$ is an \textit{embedding}. Clearly, embeddings are immersions. Hereafter, we consider $p<d$, and only assume that the map is an immersion. 

Define the tangent vectors to $\Sigma$:
\be \label{tmu}
t^\mu_i = \partial_i x^\mu ~.
\ee
Now, $M$ is a differentiable manifold endowed with a metric $g_{\mu\nu}$ that enables us to measure the distances between points.
The metric on $\Sigma$ is induced from the metric on $M$:
\be
h_{ij} = t^\mu_i t^\nu_j g_{\mu\nu} ~.
\ee
Since there are $p$ vectors tangent to the submanifold $\Sigma$, there are $d-p$ normal vectors $n^\mu_A$, $A=p+1,\ldots,d$.
At each point $Q\in\Sigma$, the tangent and normal vectors $t^\mu_i$ and $n^\mu_A$ span orthogonal subspaces.
We may choose the normal vectors to satisfy
\be
\eta_{AB} = n^\mu_A n^\nu_B g_{\mu\nu} ~, \label{eq:blah}
\ee
where $\eta_{AB}$ is a diagonal matrix with eigenvalues $\pm 1$.
As we shall soon see, the selection of a basis of normal vectors that satisfies~\eref{eq:blah} is not unique.
In fact, the normal frame will be defined only up to gauge transformations that preserve~\eref{eq:blah}.

Using $t^\mu_i$, $n^\mu_A$, $h_{ij}$, and $\eta_{AB}$, we as well decompose the inverse metric on $M$ as
\be
g^{\mu\nu} = h^{ij} t^\mu_i t^\nu_j + \eta^{AB} n^\mu_A n^\nu_B ~.
\ee
The Greek indices label the ambient space $M$.
The lowercase Latin indices label the tangent vectors, and the uppercase Latin indices label the normal vectors.
The metrics $g_{\mu\nu}$, $h_{ij}$, $\eta_{AB}$ and their inverses are used to raise and lower indices.
We can use $t^\mu_i$ and $n^\mu_A$ to trade ambient indices for tangent and normal ones.

As we traverse from point to point on the submanifold $\Sigma$, the normal vectors can of course change.
Employing the covariant derivative $\nabla_\mu$ defined using the Levi-Civita connection on $M$, we compute
\be
t^\nu_i \nabla_\nu n^{\mu A} = K_{ij}^A t^{\mu j} - T_i^{AB} n^\mu_B ~,
\ee
where $K_{ij}^A$ are the \emph{extrinsic curvatures} (or second fundamental forms) and the $T_i^{AB}$ are the \emph{extrinsic torsions}:
\bea
K_{ij}^A &=& t^\mu_i t^\nu_j \nabla_\mu n_\nu^A ~, \label{eq:ext} \\
T_i^{AB} &=& t^\mu_i n^{A \nu} \nabla_\mu n_\nu^B ~. \label{eq:tor}
\eea
Bear in mind that the extrinsic torsion is a different object from the usual torsion associated with a connection.
In what follows, as these are somewhat involved manipulations, in order to focus the conversation on the essential physics and geometry, we refer the interested reader to \ref{sec:app geometry} for further mathematical details that inform the statements that we make.


\subsection{Gauge freedom in the normal frame}\label{sec: gauge invariance}
A crucial component of the setup described in the previous section is the decomposition of the tangent bundle $TM$ on $\Sigma$.
For any point $x\in\Sigma$ vectors in $T_x M$ can be segregated into tangent components $t_i^\mu$ and normal components $n_\mu^A$.
Hereafter, we refer to the span of  $n_\mu^A$ as the \emph{normal frame}.
As a matter of fact, as shown in \ref{sec:app geometry}, this decomposition can be extended to a neighborhood of $\Sigma$.

Now, there is still an outstanding issue regarding this decomposition that we must address.
While the tangent vectors can be determined completely in terms of the immersion map~\eqref{tmu}, the normal vectors are defined indirectly via~\eqref{eq:blah} and the requirement that
\be\label{orth}
n_\mu^A t^\mu_i =0 ~. 
\ee
As we shall see, these conditions still leave some freedom in the choice of normal frame.
The most important manifestation of this freedom is the ability to choose frames with different extrinsic torsions.
In this section we provide a general discussion of this phenomenon.
The reader interested in gaining more intuition can go to Section~\ref{sec:five} where we discuss the relationship between torsion and normal frames for the familiar example of a curve in $\mathbb{R}^3$.

Let us count the number of independent components in the normal frame.
There are $d-p$ normal vectors $n_\mu^A$ with $d$ components.
Condition~\eqref{eq:blah} gives $(d-p)(d-p+1)/2$ constraints.
In turn,~\eqref{orth} fixes $p(d-p)$ components.
This leaves us with 
\be
\label{NMmodulonneta}
\mathrm{\#\ independent\;components} 
=\frac{(d-p)(d-p-1)}{2} \,.
\ee
Not coincidentally, this number matches the number of independent components of the extrinsic torsion $T_i^{AB}$ as well as the dimension of the Lie group $O(d-p)$.\footnote{
To be precise, we should take into account the signature of $M$.
Hence, if there are $k$ timelike normal directions, the group should be $O(d-p-k,k)$.
Moreover, we chose the orthonormal group because parity, \textit{i.e.}, the global change of sign for all normal vectors, is a symmetry.
In particular, for codimension one hypersurfaces, there are no $T$ and the symmetry group becomes discrete $O(1)=\mathbb{Z}_2$: the only ambiguity left is the choice of the orientation of the normal vector.}
Indeed, it is natural to think of the normal frame in the language of an $O(d-p)$ classical Yang--Mills theory living on $\Sigma$ \cite{Capovilla:2006te}.
This perspective becomes more compelling once we observe that conditions~\eqref{eq:blah} and~\eqref{orth} are still satisfied after a transformation of the form 
\be
\label{nnrotate}
n^A_\mu \to \mathcal{M}^{A}_{\;B} n^B_\mu \,,
\ee
where $ \mathcal{M}^{A}_{\;B}$ is a $\sigma^i$ dependent $O(d-p)$ matrix.

One easily sees that the extrinsic curvature transforms in the fundamental representation of $O(d-p)$, \textit{i.e.}, 
\be 
K^A_{ij} \to  \mathcal{M}^{A}_{\;B} K^B_{ij}  \,.
\ee
From this, and using the orthonormality of $\cal M$, we observe that the quantity
\be
\eta_{AB} K^A_{ij} K^B_{kl}
\label{hsquared}
\ee
is gauge invariant.
In particular, both the quadratic terms $\tr K^A K_A$ and $\tr K_A \tr K^A$ are gauge invariant, where the trace is taken over the tangent indices.
On the other hand the extrinsic torsion transforms just like a gauge field
\be
T^{AB}_i \to
{\mathcal M}^C_{\;A}{\cal M}^D_{\;B}\;T^{AB}_i + \eta^{AB} {\cal M}^{C}_{\;A}\partial_i {\cal M}^{D}_{\;B}\;.
\label{gaugeT}
\ee
Hence, we see that the extrinsic torsion transforms non-trivially as we change normal frames. Moreover, since $T^{AB}_i$ transforms like a connection we are compelled to introduce the gauge covariant derivative operator 
\be
\tilde{D}_{i\;\;B}^{\;A}V^{B}_{j\dots}
\equiv 
 \tilde{\nabla}_ iV^{A}_{j\dots}
 + 
  T_i^{AB} \eta_{BC}V^{C}_{j\dots}
\label{covariantDofK}
 \,,
\ee
to which the \textit{field strength}
\be
F_{ij}^{AB}
\equiv
\tilde{\nabla}_{[i} T_{j]}^{AB}
-
T_{[i}^{AC}  T_{j]}^{BD}
\eta_{CD} \,,
\label{fieldstrengthofT}
\ee
 can be naturally associated.

In light of these definitions, we can rewrite some of the geometric identities computed in \ref{curvature identities}.
For example, the generalized Codazzi--Mainardi~\eqref{CMlieOLD} and Ricci~\eqref{RclieOLD} equations can be recast as
\be
R^A_{\;\;j ik } =\tilde{D}_{[k\;\;B}^{\;A} K^B_{i]j}
\,,
\label{superCM}
\ee
and
\be 
F_{ij}^{AB}
=
K_{[ik}^A K_{j]l}^B h^{kl} -R^{AB}_{\;\;\;\;\;ij}
\,,
\label{superRicci}
\ee
respectively. An interesting consequence of the above equation is that only when the right hand side vanishes, is it 
possible to use gauge freedom to select - at least locally - a torsionless frame, $T^{AB}_i=0$. Observe that this is always the case for $p=1$. This prescription naturally extends to the case of any truly geometrically invariant action: it must be built using only gauge invariant quantities.
In particular, it is clear that whenever a $\tilde{\nabla}_i$ is hitting a gauge covariant quantity it has to be replaced by $\tilde{D}_{i\;\;B}^{\;A}$.  Finally, notice that~\eqref{superRicci} allows us to exchange $F_{ij}^{AB}$ for quantities on the right hand side. Therefore, for gauge invariant actions the extrinsic torsion appears only in combinations which, using~\eqref{superRicci}, can be replaced by terms depending on the extrinsic curvature and projections of the ambient curvature. 


\subsection{Dimensional analysis and the effective action}
\comment{
In quantum field theory, symmetry supplies an organizing principle.
Any terms can appear in the Lagrangian that are consistent with the symmetries of the system under consideration.
In particular, the invariant operators that describe the interactions of the theory can have very large numbers of derivatives or quantum fields.
The symmetries under which these operators are invariant can be gauge symmetries or global symmetries.
They can be continuous symmetries or discrete symmetries.

The gauge symmetry for a theory of gravity is diffeomorphism invariance, which ensures that the coordinates we assign to an event in spacetime are auxiliary labels without any intrinsic meaning.
A diffeomorphism $\varphi: M\to M$ is an active coordinate transformation.
Under this change of coordinates, the form of the equations obtained from the action remains the same, which means that the theory enjoys a general covariance.
The most general action we can write for a generally covariant theory in $d$ spacetime dimensions is
\be
S_\mathrm{grav} = \frac{1}{16\pi G_d} \int d^dx\ \sqrt{-g}\ [\Lambda + R + \alpha' ( a_1 R^2 + a_2 R_{\mu\nu} R^{\mu\nu} + a_3 R_{\mu\nu\rho\sigma} R^{\mu\nu\rho\sigma} + a_4 \nabla^2 R) + \ldots] ~, \label{eq:sgrav}
\ee
where the Newton constant $G_d$ sets the strength of the gravitational interaction.
We work in a system of units where $c=\hbar=1$ so that time has the same units as distance and length is an inverse mass.
Since the curvatures $R$, $R_{\mu\nu}$, $R_{\mu\nu\rho\sigma}$ involve two differentiations of the metric with respect to the coordinates $x^\mu$, they scale like $[\mathrm{mass}]^2$.
As the measure $d^dx\ \sqrt{-g}$ transforms as a tensor, the leading term in the action is simply proportional to the volume of spacetime.
The cosmological constant $\Lambda$ is zeroth order in the curvatures and represents the energy density of the vacuum.
At first order, the only diffeomorphism invariant quantity we can construct from the Riemann tensor is the scalar curvature $R = g^{\mu\nu} R^\lambda_{\;\;\mu\lambda\nu}$.
Taking $S_\mathrm{EH} = \int d^dx\ \sqrt{-g}\ R$ and demanding $\delta S = 0$ for variations with respect to the inverse metric $g^{\mu\nu}$, we obtain the Einstein equation of general relativity.
At the next order, we can include terms quadratic in the curvature or terms with two derivatives that act on $R$.
We introduce a new dimensionful parameter $\alpha'\propto \ell_s^2$, where $\ell_s$ is a characteristic length scale.
The $\alpha'$ prefactor then has units $[\mathrm{mass}]^{-2}$, which ensures that the second order terms in the action have the same mass dimension as the Einstein--Hilbert term $\sqrt{-g}\ R$.
The contraction of Lorentz indices ensures that the integral is a diffeomorphism invariant observable.
In principle, the coefficients $a_i$ are independent dimensionless parameters.
This continues: at cubic order, for example, we have terms like $R^3$, $R_{\mu\nu} R^\nu_{\ \rho} R^{\rho\mu}$, $(\nabla_\mu R) (\nabla^\mu R)$, etc.\ that are each accompanied by an $\alpha^{\prime2}$ prefactor and some dimensionless parameter $b_i$.

Given the immersion~\eref{eq:imm}, we want to consider the restriction of the gravitational action in the spacetime $M$ onto $\Sigma$.
We will write this up to quadratic order.
Crucially, the low energy effective action or energy functional that we construct depends purely on geometric quantities.
Applying general principles, we may write this expression purely on dimensional grounds.
}

The equations of motion which determine minimal surfaces arise from applying the variational principle to an energy functional, which we call the \textit{effective action}.
Symmetry considerations and dimensional analysis provide guiding principles in constructing the effective action.
In this work, we will keep terms up to quadratic order. Nevertheless, many of the tools developed here can be readily applied to higher order actions.

To formulate the effective action, we must first ask ourselves about the kind of terms that respect the symmetries.
The geometric functionals must satisfy certain basic requirements:
\begin{itemize}
\item To be generally covariant, the functional should depend on geometric properties of $\Sigma$ and not on specific choices of the coordinates.
This can be achieved by requiring every index to be properly contracted.
\item The formulation of the Wilsonian effective action in quantum field theory teaches us that we should organize terms in the functional according to the dimensions of their couplings.
In cases where the functional is to be interpreted as a configuration energy, higher order terms will probably contribute less to determine the local minimum, \textit{i.e.}, they would be more and more irrelevant at large wavelengths (\textit{viz.}, in the infrared). We wish to stress that this framework is used only as a guiding principle in this work. Sometimes we will take the effective action as given and not as a small deformation of other theory.

\item From the elastica perspective, the inclusion of terms up to quadratic order can be viewed as an expansion in extrinsic curvatures.
We assume that $\Sigma$ is moderately curved with respect to the microscopic scale and include only the first non-trivial contributions to the total elastic energy of the submanifold.
Higher order terms in the flat limit would vanish faster.
\item As in a standard gauge theory, we allow only gauge invariant terms in the functional under the transformation~\eqref{nnrotate}.
For example a quadratic term in the extrinsic torsions would respect the above conditions but will transform as
\be
T_i^{AB} T^{i}_{AB} \to T_i^{AB} T^{i}_{AB} + 2 T^i_{AB} \eta^{CD} \mathcal{M}^A_{\;C} \partial_i \mathcal{M}^B_{\;D} ~.
\ee
Such terms are forbidden.
Indeed, as we have noted, torsions can only appear within the field strength~\eref{fieldstrengthofT},\footnote{With the notable exception of~\eqref{tau d}.} which is a gauge invariant combination that in turn can be recast in favor of curvatures using~\eref{superRicci}.
\end{itemize}

Secondly, we consider the mass dimension of the various building blocks of the action.
We have
\bea
&& [g_{\mu\nu}]=[h_{ij}]=[\eta_{AB}]=[n_\mu^A]=[t^i_\mu]=[\mathrm{mass}]^0 ~, \nn \\
&& [K^A_{ij}] = [T^{AB}_i]=[\Gamma_{\mu\nu}^\rho]=[\tilde{\Gamma}_{ij}^k]=[\mathrm{mass}]^1 ~, \nn \\
&& [\mathcal{R}_{ijkl}]=[R_{\mu\nu\rho\sigma}] = [\mathrm{mass}]^2 ~.
\eea
We determine the dimensions of the extrinsic curvature and the torsion from inspection of~\eref{eq:ext} and~\eref{eq:tor}.
We also observe that contracting curvatures with normal and tangent vectors in order to exchange the indices does not alter the mass dimension.

With these precepts in mind, we see that we can build terms only with positive energy (and thus negative length) dimensions.
At zeroth order, the only object respecting our requirements is the identity.
This leads to an area term:
\be
S_0[\Sigma] = \lambda_0 \int_\Sigma d^p\sigma\ \sqrt{h}\ 1 = \lambda_0\, \mathrm{Area}[\Sigma] ~.
\ee
There are no terms at first order: $\tr K^A$, for example, has a free index $A$.
At second order we identify six combinations of the curvatures:
\begin{align}
S_2[\Sigma] =\int_\Sigma  d^p\sigma\ &\sqrt{h}\ \big[\lambda_1 \mathcal{R} + \lambda_2 R + \lambda_3 R_A^{\;\;\,A} +\lambda_4 R_{A B}^{\;\;\;\;\;AB}\nn\\
&+ \lambda_5 \tr K_A \tr K^A + \lambda_6 \tr\left( K^A K_A\right) \big]
\label{S2action}
\end{align}
The contracted Gauss relation~\eref{GClie2} allows us to eliminate one of these objects leaving only five independent terms.
With odd numbers of $K$s, it is not possible to simultaneously pair and contract both the tangent and the normal indexes.
Therefore, there are no terms at cubic order, and the next contribution to the energy functional arises at order four.
Schematically, these terms go like $R^2$, $RKK$, $K^4$, $\tilde D^2 R$, and $\tilde D^2 K^2$. Thus, up to second order in derivatives, we obtain the low energy action
\bea
S_\mathrm{eff}[\Sigma] = S_0[\Sigma] + S_2[\Sigma]~.
\label{eq:seff}
\eea

A final comment is in order in the special case of codimension $d-p=2$, where the gauge group is $O(2)\simeq U(1)$.
Recall that the extrinsic torsion is antisymmetric on its normal indices.
Thus, in codimension two, it is proportional to the Levi-Civita symbol $\epsilon$.
Therefore, for $p=1$ we can define the \emph{curve torsion}
\be \label{tau d}
\tau=\frac{1}{2}\epsilon_{AB} T^{AB} \,,
\ee
which transforms with a total derivative as a standard $U(1)$ gauge field.
Therefore, the integral
\be
W = \int_\Sigma \tau 
\label{twisting}
\,, 
\ee
is gauge invariant, provided fixed boundary conditions, and corresponds to the curve's twist. This term could clearly be added to the general action. However, since it is not locally gauge invariant and exists only for $d=3$ and $p=1$ we will not consider it further. Interestingly, \eqref{twisting} was introduced in the holographic entanglement entropy functional for theories dual to Topological Massive Gravity (TMG)~\cite{Castro:2014tta}.\footnote{Note that its contribution to the shape equations can be easily derived as a special case of the normal variation \eqref{Lnormalt}.}

For the case of surfaces $p=2$ we can consider instead the field strength \eqref{fieldstrengthofT}, which is antisymmetric in both normal and tangential indices. Therefore, by the same argument we can consider the term 
\be 
\varphi=\frac{1}{4}\epsilon_{AB} \epsilon^{ij}F_{ij}^{AB}\,,
\ee
which is a well-defined gauge invariant quadratic term. This term is of relevance in the study of holographic entanglement entropy for four dimensional gravitational theories with Chern--Simons terms~\cite{Azeyanagi:2015uoa, Ali:2016cng}. Notice that using the Ricci identity \eqref{superRicci} this term can be recast in terms of the extrinsic curvatures and a projection of the Riemann tensor
\be
\varphi=\frac{1}{4}\epsilon_{AB} \epsilon^{ij}\left(K_{[ik}^A K_{j]l}^B h^{kl} -R^{AB}_{\;\;\;\;\;ij}\right)\,.
\ee
Moreover, whenever $p$ is odd it is possible to define on $\Sigma$ a classical $SO(d-p)$ Chern--Simons term \cite{Chern:1974ft} which encodes topological degrees of freedom.\footnote{These terms should be distinguished from those mentioned in the previous paragraph. Gravitational Chern-Simons terms are similar to Eq.\eqref{CS} but the role of $T_k^{DC}$ is played by the spacetime's Levi-Civita connection and they are regarded as modifications to Einstein gravity.} For instance, if $p=3$ we have
\be\label{CS}
S_{CS}\sim\int_\Sigma  d^3\sigma\epsilon^{ijk}\eta_{AC}\left(F_{ij\:B}^{A} T_k^{BC}-\frac{1}{3}\,T_{i\:B}^{A} T_{j\:D}^{B}T_k^{DC}\right)\,,
\ee
which is gauge invariant up to boundary contributions. For analogous reasons to those given for \eqref{tau d} we do not consider these objects further in the present work.


\subsection{Shape equations}\label{shapeeq}
In this section, we display the equations of motion coming from extremizing the effective action~\eqref{eq:seff}. These kind of equations have been studied by a number of authors, both in the mathematics and the physics communities~\cite{10.2307/1970556, yano1978, chen1978, Erbacher, Yau, Boisseau:1992ud, hu2002willmore, Armas:2013hsa,Armas:2015kra,Forini:2015mca}. The equations presented here encompass many of these examples. They are valid for arbitrary Riemannian manifolds of any dimension and codimension, and they are gauge covariant. 
Only after deriving these equations, we became aware of works by Guven and Capovilla~\cite{PhysRevD.48.4604, PhysRevD.51.6736} as well as Carter (see \cite{Carter:1997pb} and references therein), where these results were previously derived. Nevertheless, we provide a detailed version of our derivation in \ref{sec:app geometry} and~\ref{eom}.
In terms of the notation defined in Section~\ref{notation}, the final result reads:
\be
{\cal E}^A=\lambda_0\tr K^A+\sum_{n=1}^6\,\lambda_n {\cal E}^A_n=0 ~, \label{eq:eom1}
\ee
with
\begin{align}
& {\cal E}^A_1=\tr K^A{\cal R}-2{\cal R}^{ij}K^A_{ij}, \\
& {\cal E}^A_2=\tr K^A R+ n^{A}_{\mu}\nabla^\mu R , \\
& {\cal E}^A_3=\tr K^A R_{B}^{\;\;B}+2\tilde D_k^{\,AB}R^k_{\,B}+ n_C^\mu n^{C\nu} n^{A\delta} \nabla_\delta R_{\mu\nu},\\
 &{\cal E}^A_4=\tr K^A   R_{CB}^{\;\;\;\;\;CB}+4\tilde D_k^{\,AB}R^{kC}_{\;\;\;BC}+ n_C^\mu n_B^\nu n^{C\rho} n^{B\sigma} n^{A\delta} \nabla_\delta R_{\mu\nu\rho\sigma}, \\
& {\cal E}^A_5=  \tr K_B\left[\tr K^A\tr K^B -2\tr\left(K^B  K^A\right)-2R^{B\;\;Ai}_{\;\;i}\right]\\
&\hspace{9mm}-2\tilde D_{i\;\;C}^{\;A}\tilde D^{iCB} \tr K_B,\nn \\
& {\cal E}^A_6=-  2
\left[ \tilde D_{i\;\;B}^{\;A} \tilde D_j^{\;BC} K_C^{ij}+\tr \left( K^B K_B K^A\right)+
K_B^{ij}  R^{B\;\;A}_{\;\;\;j\;\;\;i}
\right]\\
&\hspace{9mm}+\tr K^A  \tr \left( K_B K^B\right),\nn
\end{align}
where we used the covariant derivative $\tilde{D}_{i\;\;}^{\;AB}$ defined in~\eqref{covariantDofK}. In a torsionless frame, provided it exists, this covariant derivative simplifies and becomes
\be
\tilde{D}_{i\;\;}^{\;AB}\to \eta^{AB}\tilde\nabla_i\,,
\ee
which implies that the equations of motion also become simpler.
In deriving~\eqref{eq:eom1} we have made no assumptions about $\Sigma$ and $M$ beyond those stated in Section~\ref{geo setup}.
Notice that the ${\cal E}$s above are not independent, indeed, the identity 
\be
{\cal E}^A_1-{\cal E}^A_2+2{\cal E}^A_3-{\cal E}^A_4-{\cal E}^A_5+{\cal E}^A_6=0
\ee
holds. This identity can be shown by considering the normal variation of the Gauss relation \eqref{GClieOLD} and employing judiciously the second Bianchi identity and the Codazzi-Mainardi equation \eqref{CMlieOLD}.

In what follows, we shall consider a number of different cases, corresponding to a variety of applications, which give more tractable versions of~\eqref{eq:eom1}.
Hereafter, we refer to the above equations as \emph{shape equations} and to their solutions as \emph{extrema}.
The simplest examples of such extrema occur when all the coefficients in the effective action, except $\lambda_0$, vanish.
In this case, the extrema correspond to \emph{minimal submanifolds} with 
\be\label{trK0}
\tr K^A=0\,.
\ee
Familiar examples are geodesics ($p=1$) and minimal surfaces ($p=2$).


\section{Extrema in maximally symmetric spaces}\label{sec:three}
Let us consider a simplification of~\eqref{eq:eom1} that comes from restricting the ambient $M$ to a \emph{maximally symmetric space} (MSS). For the moment, we leave the dimension $d$ and codimension $d-p$ arbitrary.
Later, we shall consider some cases that lead to further simplifications. For a maximally symmetric space, the Riemann curvature tensor can be written as 
\be
R_{\mu\nu\rho\sigma} = \frac{R}{d(d-1)} \left(g_{\mu\rho} g_{\nu\sigma} - g_{\mu\sigma} g_{\nu \rho}\right) ~, \label{MSPRie}
\ee
where the scalar curvature $R$ is a constant.
The Ricci tensor then reads
\be
R_{\mu\nu} = \frac{R}{d} g_{\mu\nu} ~,
\ee
and the geometry enjoys $\frac12d(d+1)$ Killing directions corresponding to a maximum number of isometries.
The normal projections are
\begin{align}
&R_{ABCD}=\frac{R}{d(d-1)} \left(\eta_{AC} \eta_{BD} - \eta_{A D} \eta_{BC}\right) ~, \\
&R^{\,\;BCA}_{C} = \frac{d-p-1}{d(d-1)}\, R\, \eta^{AB} ~, \\
&R_{AB} =  \frac{R}{d} \eta_{AB}
\end{align}
whose contractions are readily calculated:
\begin{align}
&R_A^{\;\;\,A}=\frac{(d-p)}{d} R ~, \\
&R_{A B}^{\;\;\;\;\;AB}=\frac{(d-p-1)(d-p)}{d(d-1)} R ~.
\end{align}

With the above identities we can simplify the effective action and find
\be
S_\mathrm{eff}[\Sigma]= \int_\Sigma d^{p}\sigma\ \sqrt{h}\ \left[\hat\lambda_0+\lambda_1{\cal R}+\lambda_5\tr K_A \tr K^A+\lambda_6 \tr (K^A K_A)\right] ~,
\label{eq:qwe}
\ee
with
\be\label{lambda 0 e}
\hat\lambda_0=\lambda_0+\frac{\kappa}{L^2}\left[\lambda_2 d(d-1)+\lambda_3(d-1)(d-p)+\lambda_4(d-p-1)(d-p)\right] ~,
\ee
and the radius of curvature $L$ is defined via the expression
\be
R=\kappa\frac{d(d-1)}{L^2} ~, \qquad \kappa=0,\pm 1 ~.
\ee
The terms in the effective action~\eref{eq:qwe} are not all independent. Indeed,
in the present context the contracted Gauss identity~\eqref{GClie2} is given by
\be
\mathcal{R} = \frac{\kappa\, p(p-1)}{L^2} - \tr \left(K^A K_A\right) + \tr K^A \tr K_A ~.
\ee
With this identity we can always trade one of the curvature invariants in~\eqref{eq:qwe}.
For instance, we can write
\begin{align}\label{seff Max sym}
S_\mathrm{eff}[\Sigma]=
\int_\Sigma d^{p}\sigma\ \sqrt{h}\ & \big[ (\hat\lambda_0+\hat\lambda_6 p(p-1))+(\lambda_1-\lambda_6) {\cal R}\nn\\
&+(\lambda_5+\lambda_6)\tr (K_A) \tr( K^A)\big]\, ,
\end{align}
where $\hat\lambda_i = \frac{\kappa}{L^2}\lambda_i$, for $i=1,5,6$.
Which curvature term we choose to eliminate is a matter of convenience.

From the functional~\eqref{seff Max sym}, equation~\eqref{eq:eom1} reduces to 
\begin{align}\label{shape mss}
0=&(\hat\lambda_0+\hat\lambda_6 p(p-1))\tr K^A+(\lambda_1-\lambda_6)\left( \tr K^A{\cal R}-2{\cal R}^{ij}K^A_{ij}
 \right)\nn\\
&-2(\lambda_5+\lambda_6) \tilde D^{iAC}\tilde D_{iCB}\tr K^B+(\lambda_5+\lambda_6)\tr K^A  \tr K_B\tr K^B\nn\\
&-2(\lambda_5+\lambda_6) \tr K_B
\Big[ \tr\left( K^B K^A\right)
+
p\frac{\kappa}{L^2}\eta^{AB}\Big]\,.
\end{align}
An interesting consequence of this equation is that, in maximally symmetric spaces, minimal submanifolds~\eqref{trK0} are extrema of the full functional~\eqref{eq:seff} if either
\be\label{cond minim}
\lambda_1=\lambda_6\qquad\text{or}\qquad {\cal R}^{ij}K^A_{ij}=0 ~.
\ee
The fulfillment of the first condition will depend on the physics being considered. Notice that the second condition is always satisfied for curves and surfaces ($p=1,2$).
Indeed, for $p=1$ the intrinsic geometry is trivial while for $p=2$:
\be
{\cal R}^{ij}K^A_{ij}=\frac{{\cal R}}{2}\tr K^A~.
\ee 
On the other hand, for $p>2$, minimal submanifolds do not necessarily satisfy the shape equations.

\subsection{Curves in maximally symmetric surfaces}\label{sec:curves}
Now, we wish to go beyond minimal submanifolds and study other classes of extrema.
In the following, we restrict to a simple, yet rich, example.
These are curves in maximally symmetric surfaces 
(\textit{i.e.}, $d=2$, $p=1$).
Here, the frame is automatically torsionless, and there is only a single non-vanishing extrinsic curvature, which we denote by $k$.
The relevant functional reads
\bea
\label{mss curve}
S_\mathrm{eff}[\Sigma]=
\int_\Sigma d^{p}\sigma\ \sqrt{h}\ \left[\hat\lambda_0+\lambda'_5\tr (k)^2\right] \,,
\eea
where $\hat\lambda_0$ is given by~\eqref{lambda 0 e} and $\lambda_5'= \lambda_5+\lambda_6$.
Thus, the shape equation~\eqref{eq:eom1} becomes
\be\label{Curve in MSS}
2\tilde\Delta \tr k+\tr k^3-\left(\frac{\hat\lambda_0}{\lambda'_5}-\frac{2\kappa}{L^2}\right) \tr k=0 ~.
\ee
If we parameterize the curve by its arclength $s$ measured in units of $L$, then $h=1$ and~\eqref{Curve in MSS} reads
\be\label{Curve in MSS arc}
2\ddot k+ k^3- B\, k = 0 ~, \qquad B = \left(\frac{\hat\lambda_0}{\lambda'_5}-\frac{2\kappa}{L^2}\right) ~,
\ee
where $\dot\;=d/ds$.
Indeed, geodesics $k=0$ solve the above equation as discussed before.
The first kind of non-geodesic solutions of~\eqref{Curve in MSS arc} are
\be\label{CMC}
k^2 = B = \mathrm{constant} ~,
\ee
which are constant mean curvature (CMC) solutions.
Clearly, these solutions exist provided $B>0$ which imposes a bound that relates the coupling constants in the action and the curvature of the ambient space
\be\label{bound1}
\frac{\hat\lambda_0}{\lambda_5}>\frac{2\kappa}{L^2}\,.
\ee
We will return to these solutions in Section~\ref{EE 3d}.
Interestingly enough, the differential equation \eqref{Curve in MSS arc2} is formally equivalent to the equation of motion of a classical field in an quartic potential unbounded from below
\be 
V(k) = \frac{1}{8} k^2 (2 B-k^2) \,.
\label{curvaturepotential}
\ee
For $B>0$, this potential has two maxima at $k=\pm \sqrt{B}$ and a local minimum at $k=0$; meanwhile,  for $B\leq 0$, $k=0$ is the only maximum. Notice that these extrema correspond to the constant mean curvature and geodesic solutions, respectively.

\begin{figure}[h]
\centering
\includegraphics[scale=0.29]{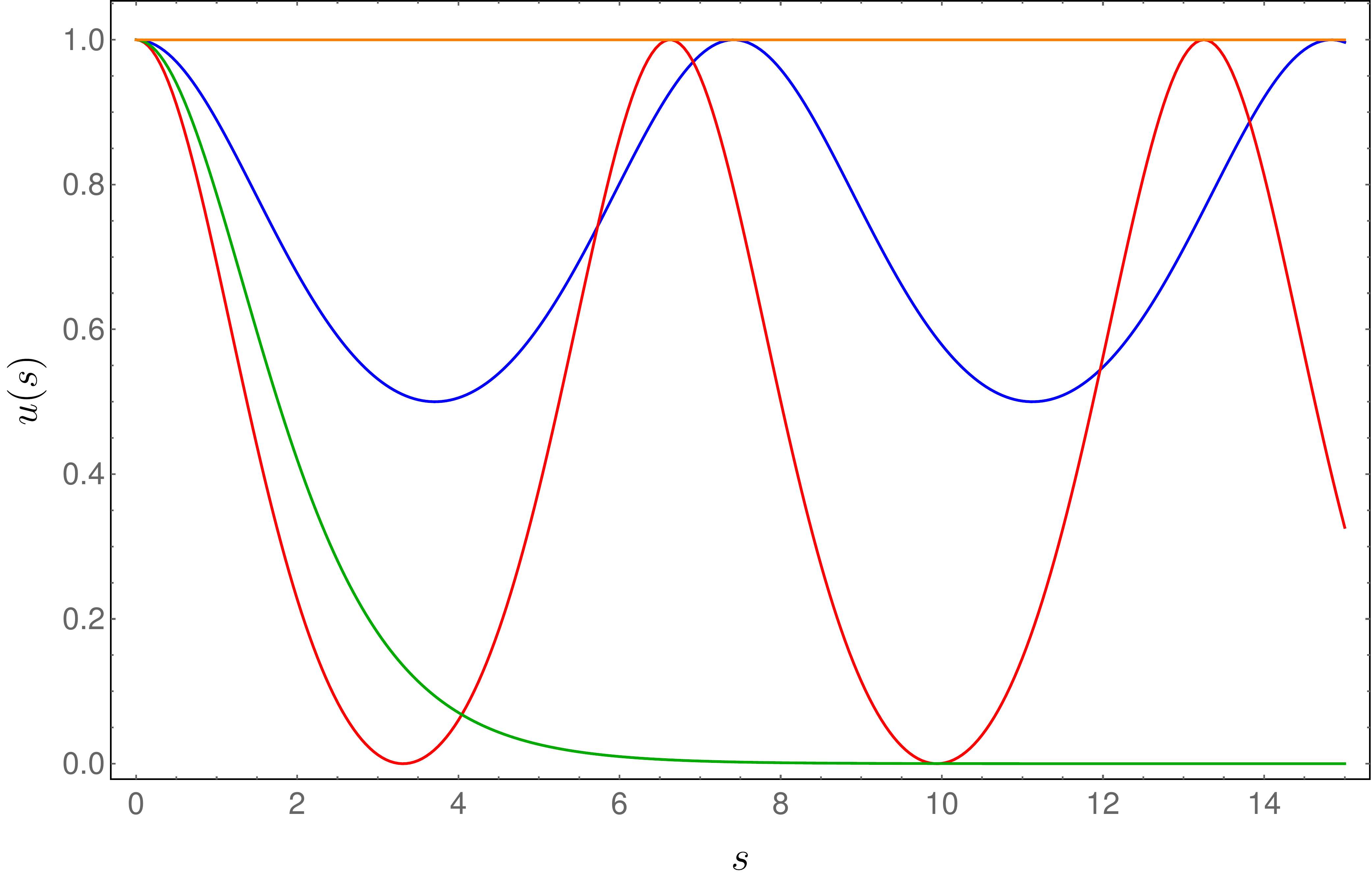}
\caption{Behavior of the extrinsic curvatures, $u(s)=k^2(s)$, for extrema in maximally symmetric spaces. The orange curve corresponds to a CMC  \eqref{CMC}, the red one is wavelike \eqref{op1}, the blue curve is orbitlike \eqref{op2} and the green one is asymptotically geodesic \eqref{op3}.}
\label{fig:u}
\end{figure}

As explored previously in~\cite{Langer1984}, it is possible to find solutions with non-constant mean curvature analytically.
We proceed as follows, we multiply~\eref{Curve in MSS arc} by $\dot{k}\ne 0$ and set $u=k^2$.
Integrating, we then find an equation of form
\begin{align}\label{Curve in MSS arc2}
\dot u^2=-(u-\alpha)(u-\beta)(u-\gamma) ~.
\end{align}
The general solution to~\eref{Curve in MSS arc2} is 
\be
u(s) =k^2(s)= \alpha \left[1-\frac{\alpha-\gamma}{\alpha} \sn^2(\frac12\sqrt{\alpha-\beta}\, s,\,\frac{\alpha-\gamma}{\alpha-\beta}) \right] ~.
\label{eq:u-of-s}
\ee
(See \ref{sec:jac} for a brief recapitulation of Jacobi elliptic functions such as $\sn(z,m)$, $\cn(z,m)$, and $\dn(z,m)$.)
Using elliptic function identities, this solution enjoys a symmetry under permutation of the roots.
The second argument of the elliptic function is the elliptic modulus $m$.
We adopt the convention that the elliptic modulus $0< m< 1$ in writing our solutions explicitly.
Introducing the notation 
\be
B_\pm = B\pm \sqrt{B^2+A}\,,
\ee
where $A$ is an integration constant, the roots $\alpha$, $\beta$, and $\gamma$ for the present case are $B_\pm$ or zero.
Non-trivial solutions arise from choosing $\alpha=B_+$.
\begin{itemize}
\item Setting $\gamma=0$, the solution~\eqref{eq:u-of-s} becomes
\be\label{op1}
u(s) =B_+ \cn^2\Big(\frac12\sqrt{B_+-B_-}\, s,\,\frac{B_+}{B_+-B_-}\Big)  ~.
\ee
This form of the solution corresponds to positive $A$ so that $B_+ \ge 0 \ge B_-$.
\item Setting $\beta=0$, the solution~\eqref{eq:u-of-s} becomes
\be\label{op2}
u(s) = B_+  \dn^2\Big(\frac12\sqrt{B_+}\, s,\,\frac{B_+-B_-}{B_+}\Big)  ~.
\ee
Here, $A$ is negative so that $B_+ \ge B_- \ge 0$.
Indeed, as $\cn(\sqrt{m}\, z, m^{-1}) = \dn(z,m)$, the expressions~\eqref{op1} and~\eqref{op2} are formally the same.
We simply require that the elliptic modulus $0<m<1$ in determining which form of the solution to use.
\item If $B_-=0$, then $A=0$.
The two previous cases coincide in this case.
We have the limit $m\to 1$ of the expressions~\eqref{op1} and~\eqref{op2}.
The solution is
\be\label{op3}
u(s) =2 B \, \sech^2\Big(\sqrt{\frac{B}{2}}\, s\Big) ~.
\ee
\end{itemize}
The three solutions are, respectively, called \emph{wavelike}, \emph{orbitlike}, and \emph{asymptotically geodesic} in \cite{Langer1984}.
When $\beta=0$, we have seen that $A$ is negative.
Demanding that the roots remain real, $A$ cannot become too negative.
If $B_+=B_-$ (\textit{i.e.}, $A=-B^2$), we return to the constant mean curvature solutions for which $u(s) = B$.
The qualitative behavior of the extrinsic curvatures is different in each of the regimes as we show in Figure~\ref{fig:u}. 

We have computed the extrinsic curvature, and it is possible to use this to calculate the on-shell value of the effective action.
Substituting~\eqref{op2}, we have
\bea
S_\mathrm{eff}^\mathrm{on-shell}[\Sigma] &=& \int_0^{\ell_\Sigma} ds \left[ \hat\lambda_0 + \lambda'_5 u(s) \right] \nn \\
&=& \hat\lambda_0 \,\ell_\Sigma + 2\lambda'_5 \sqrt{B_+} E\Big(\am\big(\frac{\sqrt{B_+}}{2}\,\ell_\Sigma,m\big),m\Big) ~,
\eea
where $\ell_\Sigma$ is the total length of $\Sigma$ and 
\be
m=\frac{B_+-B_-}{B_+} ~.
\ee
Similarly, using~\eqref{op1}, we derive
\begin{align}
S_\mathrm{eff}^\mathrm{on-shell}[\Sigma] &= \hat\lambda_0 \,\ell_\Sigma + B_+ (1-m^{-1}) \,\ell_\Sigma \\
& + 2\sqrt{B_+-B_-}\, E\Big(\am\big(\frac{\sqrt{B_+-B_-}}{2}\,\ell_\Sigma,m\big),m\Big) ~,\nn
\end{align}
with
\be
m=\frac{B_+}{B_+-B_-} ~.
\ee
We expressed these results in terms of the Jacobi amplitude \eref{eq:amp} and the incomplete elliptic integral of the second kind \eqref{incom}.


\section{Holographic entanglement entropy}\label{sec:four}

Entanglement is one of the most profound and engaging aspects of quantum mechanics.
Essentially, it consists of the fact that even when we possess a complete description of a quantum system, this does not imply that we can describe every possible subsystem in a complete fashion.
The \emph{entanglement entropy} (EE) of a subsystem is a quantitative embodiment of this phenomenon.
The entanglement entropy is defined as follows.
Let $\rho$ be the density matrix of the whole system and suppose that the Hilbert space ${\cal H}$ can be factorized as ${\cal H}={\cal H}_A\otimes {\cal H}_{A^c}$, where $A$ labels the subsystem of interest and $A^c$ its complement.
We may regard $A$ as a system and $A^c$ as the environment with which the system interacts.\footnote{
In our discussions $A$ will correspond to a region in space.}
Then, by tracing over the Hilbert space of the complement, we may construct the \emph{reduced density matrix} $\rho_A=\tr_{{\cal H}_{A^c}}\rho$.
The entanglement entropy of $A$ is the Von Neumann entropy of $\rho_A$, which is
\be
S_\mathrm{EE}(A)=-\tr \rho_A\log\rho_A \,.
\ee
This notion can be defined for quantum field theories if one proceeds carefully, and it is found that the entanglement entropy encodes physics within its divergent structure.
Computations of entanglement entropy, in general, can be rather difficult especially in higher dimensions.
However, there is a great body of literature with many results, both analytical and numerical; see, for example~\cite{Horodecki:2009zz, Rangamani:2016dms} and references therein.

During the past decade, entanglement entropy has been the subject of intense study.
This is in great part due to the reformulation of the problem, under the light of the AdS/CFT correspondence \cite{Maldacena1998}, by Ryu and Takayanagi (RT)~\cite{Ryu2006}.
This proposal has been used with great success to investigate a wide variety of systems.
In its original form, the Ryu--Takayanagi prescription states that for a theory with an Einstein gravity dual, the computation of the entanglement entropy can be recast as a minimal submanifold problem in an asymptotically AdS (AAdS) spacetime.
From a practical standpoint, in order to compute the entanglement entropy for a subsystem $A$ in the boundary theory, one needs to extremize the functional
\be
S_\mathrm{eff}[\Sigma]=\frac{1}{4G_d}\int_{\Sigma} d^p \sigma\,\sqrt{h}
\ee
in an AAdS ambient space $M$, where $\Sigma$ is codimension two, is anchored at $\partial A$ and $G_d$ is the $d$ dimensional Newton's constant.
It is clear that this functional corresponds to~\eqref{eq:seff}, where the only non-vanishing coefficient is 
\be\label{l0}
\lambda_0=\frac{1}{4G_d}\,.
\ee
Therefore, the equation of motion relevant for this problem is 
\be
\tr K^A=0\,,
\ee
and the Ryu--Takayanagi prescription says that
\be 
S_\mathrm{EE}(A)=S_\mathrm{eff}^\mathrm{on-shell}[\Sigma]\,.
\ee

The Ryu--Takayanagi prescription is valid for field theories whose holographic dual can be described using Einstein gravity.
However, we know that Einstein gravity can receive higher derivative corrections, which in the context of string theory can be viewed as the result of an $\alpha'$ expansion.
The question of whether the Ryu--Takayanagi prescription is suitable in the presence of these additional terms has been explored in a number of papers~\cite{Fursaev2013, Bhattacharyya2014, Camps2014} culminating with a general prescription presented in~\cite{Dong2014}.
As it turns out, the Ryu--Takayanagi functional must be modified in a non-trivial manner;
for example, for a four derivative gravity theory with Lagrangian
\be\label{HC}
{\cal L}=-2\Lambda + R+c_1 R^2+c_2 R_{\mu\nu} R^{\mu\nu}+c_3 R_{\mu\nu\rho\sigma} R^{\mu\nu\rho\sigma}\,,
\ee
the functional that provides the entanglement entropy reads
\begin{align}\label{Dong}
S_\mathrm{eff}=\frac{1}{4G_d}\int_{\Sigma} d^p \sigma\sqrt{h} \Big[&1+2c_1 R+c_2\left(R_A^{\;\;A}-\frac{1}{2}\tr K_A\tr K^A\right)\nn\\&+2c_3\left(R_{A B}^{\;\;\;\;\;AB}- \tr (K^A K_A)\right)\Big]\,,
\end{align}
where the ambient manifold is AAdS. The question of which surface must be plugged into this functional to obtain the right value for the entanglement entropy remains open. A natural conjecture was proposed in~\cite{Dong2014} whereby the surface in question is obtained from minimizing the functional~\eqref{HC}. Indeed, in that work it was shown that for functionals of the form~\eqref{Dong}, the equations of motion match those emerging from the procedure outlined in~\cite{Lewkowycz:2013nqa}. However, as the equations of motion give rise to many possible solutions, determining which of these solutions is the one that yields the correct value of the entanglement entropy is not settled. Investigations in this direction appear in, for example,~\cite{Bhattacharyya:2014yga, Erdmenger2014,Hosseini2015, Ghodsi:2015gna}.  Clearly, the functional \eqref{HC} is of the form~\eqref{eq:seff}.\footnote{ With coefficients: $\lambda_0$ as in~\eqref{l0}, $ \lambda_1=0$, $ \lambda_2=2c_1 \lambda_0 $,  $\lambda_3=c_2\lambda_0$, $\lambda_4=2c_3\lambda_0$, $2 \lambda_5=-c_2 \lambda_0$ and $\lambda_6=-2c_3\lambda_0$.} Thus, the equations of motion are a special case of the shape equations~\eqref{eq:eom1}. There is an important point that we wish to stress: in the following sections we will regard \eqref{Dong} as a definition of the action and not in a Wilsonian spirit. We will use this functional to compute entanglement for duals to New Massive Gravity, where the deformation parameter (the inverse graviton's mass) is not small.

 The geometric perspective presented here was overlooked in the aforementioned works. There, a parametrization was proposed for the entangling surfaces leading to fourth order, highly nonlinear, differential equations. The advantage of using the shape equations~\eqref{eq:eom1} is that they display a more transparent structure. For example, at least for maximally symmetric spaces, they allow for hierarchical approach to the solution. Namely, one can solve first a second order differential equation for the extrinsic curvatures and afterwards extract the entangling surface from the extrinsic curvatures. In the following, we use this strategy and find, analytically, all the possible entangling curves for gravitational theories of the form \eqref{HC} in AdS$_3$.

\subsection{Entanglement from three dimensional gravity}\label{EE 3d}
In this section, we study the entanglement entropy for two dimensional conformal field theories (CFT$_2$) whose dual is a gravitational theory in three dimensions with a Lagrangian of the form~\eqref{HC}.
For most of the discussion below we will keep the coefficients $c_i$ arbitrary and only later commit to a particular higher derivative theory.
The only assumption we need for now is that the theory in question admits an AdS$_3$ background
\be
ds^2=\frac{L^2}{z^2}\left(-dt^2+dx^2+dz^2\right)\,.
\ee
To compute the entanglement entropy for an interval $A=\left[-\ell/2,\ell/2\right]$ in a CFT$_2$ holographically,
we consider a constant time slice of AdS$_3$, that is, a two dimensional Lobachevsky space $\mathbb{H}^2$.
Thus, the higher curvature entanglement entropy functional~\eqref{Dong} reduces to~\eqref{mss curve}. 

As discussed in Section~\ref{sec:three} the simplest extrema of this functional are geodesics, \textit{i.e.}, curves with $\tr\, k=0$.
The extrinsic curvature in $\mathbb{H}^2$ is given by~\eqref{k uhp}.
Furthermore, we are interested in a geodesic that meets the boundary at the endpoints of the interval $A$.
Demanding this, we find the curve
\be\label{geo}
z^2(s)+x^2(s)=\left(\frac{\ell}{2}\right)^2\,,
\ee
which indeed has vanishing extrinsic curvature.
The on-shell value of the functional is divergent, and this leading divergence reads as
\be\label{SGeo}
S_\mathrm{eff}^\mathrm{Geo}[\Sigma]=\hat\lambda_0\int_{\Sigma} d s=2\hat\lambda_0L\log\left(\frac{\ell}{\epsilon}\right)+{\cal O}(\epsilon)\,,
\ee
where $\epsilon>0$ is an ultraviolet cutoff.

We learned in Section~\ref{sec:curves} that there are other kinds of extrema for curves in maximally symmetric spaces, such as $\mathbb{H}^2$, besides the geodesics.
First, we turn our attention to the constant mean curvature solutions,~\eqref{CMC}, which for $\mathbb{H}^2$ obey
\be\label{kB}
k^2=B= \left(\frac{\hat\lambda_0}{\lambda_5}+\frac{2}{L^2}\right)\,.
\ee
Once more, we wish to find curves that meet the boundary at the endpoints of the interval $A$.
We find that the two solutions
\be\label{CMC hyperbolic}
x^2(s)+\left[z(s)- \left(\frac{\ell}{2}\right)\eta\right]^2=\left(\frac{\ell}{2}\right)^2\left(1+\eta^2\right)\qquad \eta=\pm\frac{L|k|}{\sqrt{1-L^2k^2}}\,
\ee
satisfy these conditions. Observe that the curves \eqref{CMC hyperbolic} exist provided that
\be\label{bound2}
k^2< \frac{1}{L^2}\,.
\ee
This last statement is a general feature of constant mean curvature solutions in hyperbolic space.
 Note that these solutions correspond to those found in \cite{Ghodsi:2015gna}. Finally, combining \eqref{bound1} and \eqref{bound2} we find that the solutions \eqref{kB} exist only if
\be\label{Willmore bound}
-\frac{2}{L^2}<\frac{\hat \lambda_0}{ \lambda_5}  < -\frac{1}{L^2}\, .
\ee
 Plugging~\eqref{CMC hyperbolic} back into the functional~\eqref{mss curve} we get the on-shell value
\begin{align}\label{SCMC}
S_\mathrm{on-shell}^\mathrm{CMC}[\Sigma]
= \frac{4}{L}\sqrt{-\lambda_5(\lambda_5+L^2\hat\lambda_0)}\log\left(\frac{\ell}{\epsilon}\right)+{\cal O}(\epsilon)\,.
\end{align}
There are other classes of extrema that can be anchored at the endpoints of $A$ in $\mathbb{H}^2$, namely, the wavelike~\eqref{op1} and the asymptotically geodesic~\eqref{op3} solutions. The latter solution has the same ultraviolet behavior as the geodesic solution, and hence, it has the same leading divergence for the on-shell value of the functional. On the other hand, the former leads to a different value altogether.

Finding the wavy solutions explicitly is significantly more complicated, and it is done in \ref{sec:wavy}.
The  arclength parametrization of these extrema can be found in equation~\eqref{wavy explicit}.
For these solutions the leading divergence of the on-shell value of~\eqref{mss curve} reads
\be\label{SWavy}
S_\mathrm{on-shell}^\mathrm{Wavy}[\Sigma]=2\hat\lambda_0 \,\ell_\Sigma+\lambda_5'\left(2-C+\frac{2C\, E\left(\frac{2+C+\lambda}{2C}\right)}{K\left(\frac{2+C+\lambda}{2C}\right)}\right)\,\ell_\Sigma+\dots
\ee
where $\lambda=\hat\lambda_0/\lambda_5'$, $C=\sqrt{A+(2+\lambda)^2}$ and $\ell_\Sigma$ is the regularized arclength of the wavelike extremum $\Sigma$, which is given by
\be
\ell_\Sigma={\cal P}\log\left(\frac{\ell}{\epsilon}\right)+{\cal O}(\epsilon)\,,
\ee
with
\be\label{P wavy}
{\cal P}=\frac{-8(C+2) K\left(\frac{2+C+\lambda}{2 C}\right)}{{\cal A}\left[\left(C-2\right)\Pi\left(\frac{4(2+C+\lambda)}{(2+C)^2},\frac{2+C+\lambda}{ 2 C}\right)-(2+C)K\left(\frac{2+C+\lambda}{2 C}\right)\right]}\,,
\ee
where ${\cal A}$ is given by~\eqref{calA}, and $K$, $E$, and $\Pi$ are complete elliptic integrals of the first, second, and third kind, respectively. See \ref{sec:jac} for details.

\begin{figure}[h]
\centering
\includegraphics[scale=0.3]{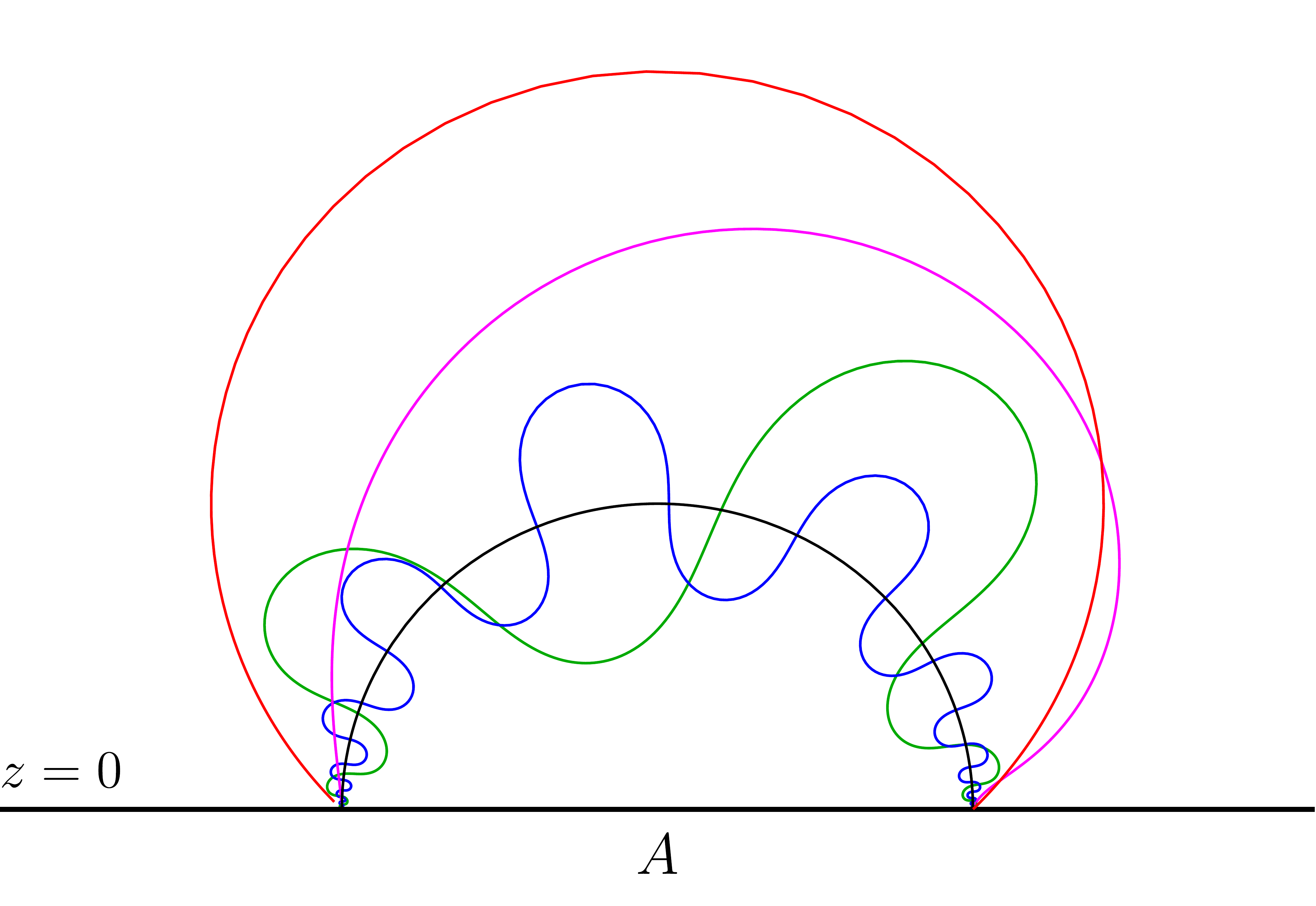}
\caption{Extrema in a constant time slice of AdS$_3$ anchored at the ends of the interval $A$. For this plot we take $L=1$ and  $\lambda=-3/2$. The black curve corresponds to the geodesic solution \eqref{geo}, the red curve is a CMCs \eqref{CMC hyperbolic}, the green and blue curves are examples of wavelike solutions \eqref{op1} with $A=50$ and $A=1000$ respectively, and the magenta solution is an asymptotically geodesic curve \eqref{op3}. Bear in mind that every type of solution is unique up to isometries that leave the interval $A$ invariant. }
\label{fig:wu}
\end{figure}

Before proceeding to a systematic comparison of the on-shell values for the different extrema, let us make one general observation.
The single interval entanglement entropy for a CFT$_2$ is given by~\cite{Holzhey1994}
\be\label{EE CFT}
S_\mathrm{EE}(A)=\frac{c}{3}\log\left(\frac{\ell}{\epsilon}\right)+{\cal O}(\epsilon)\,,
\ee
where $c$ is the central charge of the CFT$_2$.
For any parity preserving theory of higher derivative gravity admitting an AdS$_3$ background, the central charge of the dual theory can be found using the formula~\cite{Saida2000}
\be\label{central charge}
 c=\frac{L}{2G_3}\,g^{\mu\nu}\frac{\partial {\cal L}}{\partial R_{\mu\nu}}\,,
\ee
which, in the Einstein gravity limit, reduces to the Brown--Henneaux central charge~\cite{Brown1986}
\be
c_\mathrm{BH}=\frac{3L}{2G_3}\,.
\ee
For a theory with Lagrangian~\eqref{HC}, we find from~\eqref{central charge} that
\be
c=c_\mathrm{BH}-\frac{6}{LG_3}\left(3c_1+c_2\right)\,,
\ee
which implies that $c=6\hat\lambda_0L$.
Thus, we find
\be\label{equa}
S_\mathrm{EE}(A)=S_\mathrm{on-shell}^\mathrm{Geo}[\Sigma]\,,
\ee
which proves that regardless of the explicit coefficients of the Lagrangian~\eqref{HC}, the geodesics are the extrema that provide the correct value for the entanglement entropy.

Now, we address the question of minimality.
For concreteness, we will compare the on-shell values for the geodesic~\eqref{SGeo}, the constant mean curvature~\eqref{SCMC}, and the wavelike solution~\eqref{SWavy} for a specific higher curvature theory of gravity in three dimensions. For related work see~\cite{Erdmenger2014, Hosseini2015,MohammadiMozaffar2016,Ghodrati:2016ggy}.
By a simple counting argument one can show that a massless graviton in three dimensions cannot have propagating degrees of freedom.
This feature makes three dimensional gravity more tractable from an analytic point of view~\cite{Carlip:2005zn}.
By contrast, a massive graviton in three dimensions will carry two propagating degrees of freedom and allows for more complicated dynamics.
A diffeomorphism and parity invariant theory of three dimensional gravity was constructed in~\cite{Bergshoeff:2009hq}.
It is known as New Massive Gravity (NMG), and its Lagrangian reads
\be
{\cal L}_\mathrm{NMG}=-2\Lambda + R+\frac{1}{m^2}\left(R_{\mu\nu}R^{\mu\nu}-\frac{3}{8}R^2\right)\,,
\ee
where $m$ is the graviton's mass.
The coefficients of the entanglement entropy functional for New Massive Gravity in AdS$_3$ are
\be\label{lambda NMG}
\hat\lambda_0=\frac{1}{4G_3}\left(1+\frac{1}{2L^2m^2}\right)\qquad \lambda_5=-\frac{1}{8 m^2 G_3} \,.
\ee
The on-shell value for the geodesic~\eqref{SGeo} becomes
\be
S_\mathrm{on-shell}^\mathrm{Geo}[\Sigma]=\frac{L}{2G_3}\left(1+\frac{1}{2L^2m^2}\right)\log\left(\frac{\ell}{\epsilon}\right)+{\cal O}(\epsilon)\,,
\ee
and for the constant mean curvature solution~\eqref{SCMC}
\be 
S_\mathrm{on-shell}^\mathrm{CMC}[\Sigma]=\frac{1}{\sqrt{2}G_3 m}\log\left(\frac{\ell}{\epsilon}\right)+{\cal O}(\epsilon)\,.
\ee
Observe that for New Massive Gravity, the bound \eqref{Willmore bound} on the existence of constant mean curvature extrema reads
\be\label{Willmore NMG}
0\leq m^2\leq \frac{1}{2L^2}\,.
\ee
The corresponding expression for the on-shell values of the wavelike solutions is not particularly illuminating but can easily be obtained from substituting the couplings~\eqref{lambda NMG} into~\eqref{SWavy} and~\eqref{P wavy}.

\begin{figure}[h!]
\centering
\includegraphics[scale=0.36]{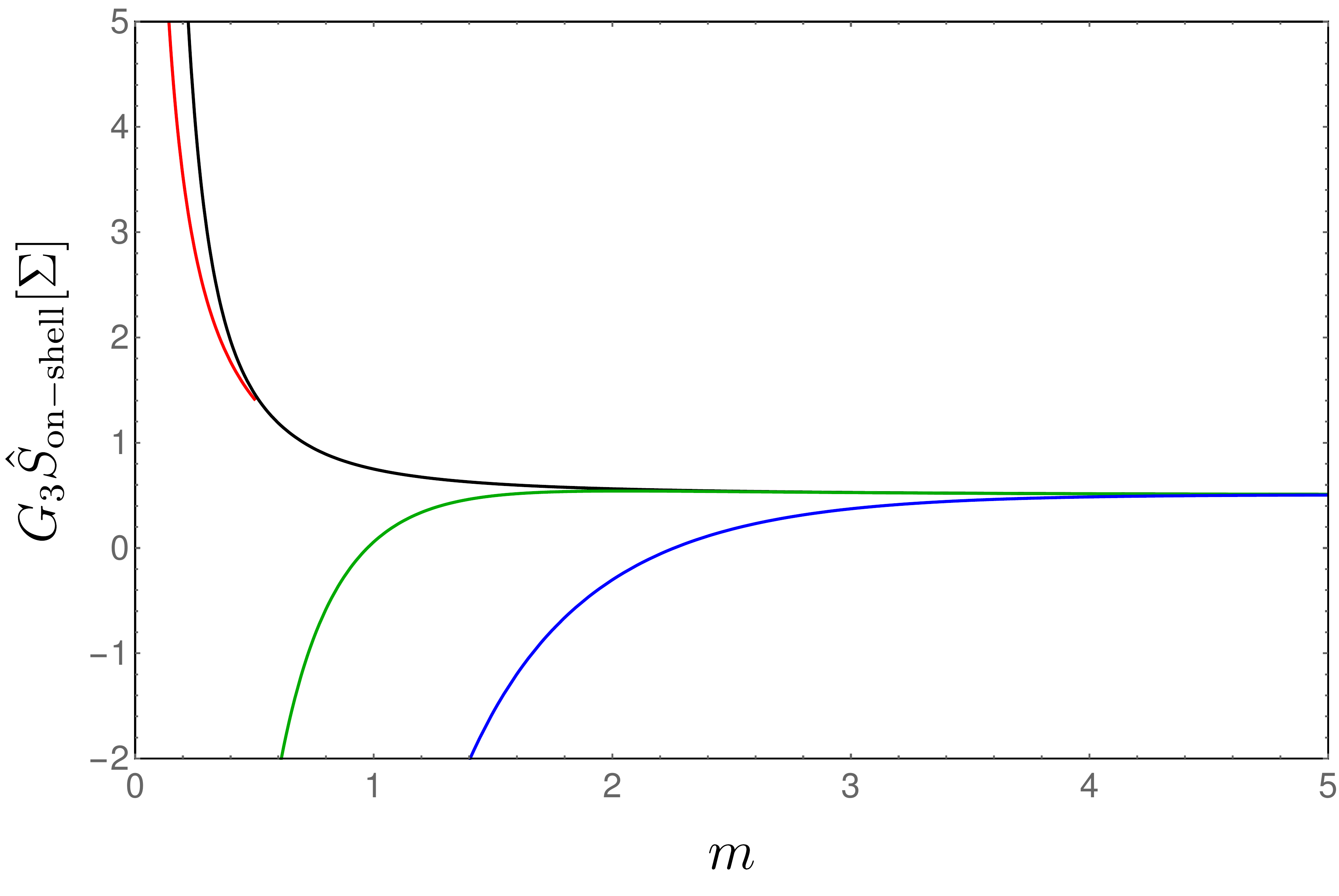}
\caption{Comparison of the renormalized on-shell values Eq.\,\eqref{renor} for the curves depicted in Fig.\,\ref{fig:u}. We keep the same values for the parameters and coloring code as in Fig.\,\ref{fig:u} i.e. black for the geodesic, red for the CMC and so on. The asymptotically geodesic curve (magenta) is absent since its renormalized on-shell value is identical to that of the geodesic. Notice that the geodesic value is always the largest.  }
\label{fig:wavy vs geo}
\end{figure}

We wish to compare the universal parts of these quantities, which can be extracted using 
\be\label{renor}
\hat S_\mathrm{on-shell}[\Sigma]=\ell\frac{d}{d\ell}\, S_\mathrm{on-shell}[\Sigma]\,.
\ee
Geodesic and constant mean curvature results can be easily compared, and we obtain 
\be
\frac{\hat S^{\text{Geo}}_{\text{on-shell}}}{\hat S^{\text{CMC}}_{\text{on-shell}}}=\frac{1+2(mL)^2}{2\sqrt{2} mL}\geq 1\, .
\ee
We find that the on-shell value of the functional is smaller for the constant mean curvature curve (whenever its existence is allowed by the bound~\eqref{Willmore NMG}), consistent with the results presented in~\cite{Ghodsi:2015gna}. Moreover, we find that the on-shell values for the wavelike solutions~\eqref{SWavy} are also smaller than those corresponding to the geodesic. See Figure~\ref{fig:wavy vs geo} for a comparison of the different extrema depicted in Figure~\ref{fig:u}. Therefore, the geodesics do not constitute a global minimum. Nevertheless, as seen in~\eqref{equa}, they always provide the correct value for the entanglement entropy.

\subsection{Holographic entanglement for logarithmic CFT}

In this section we briefly discuss the functional~\eqref{eq:seff} for New Massive Gravity backgrounds, which are conjectured to be dual to logarithmic conformal field theories (LCFT)~\cite{Gurarie1993}. These kind of theories have a wide range of applications, which include topics such as percolation, quenched disorder, and self avoiding walks. See~\cite{Hogervorst:2016itc} for a modern perspective on the subject. The single interval entanglement entropy for LCFTs has been studied from a holographic point of view~\cite{Alishahiha2014} as well as with more direct methods~\cite{bianchini2014}. Here, we revisit the computation presented in~\cite{Alishahiha2014} and find some discrepancies.

The line element dual to the LCFT reads \cite{Alishahiha2011}
\be\label{log metric}
ds^2=\frac{L^2}{z^2}\left[dz^2-2dx_+dx_- -\beta\log\left(\frac{z}{L}\right)dx_+^2\right]\,,
\ee
where we define light--cone coordinates through
\be 
t = \frac{1}{\sqrt{2}} (x_+ + x_-) \;,\quad
x = \frac{1}{\sqrt{2}} (x_+ - x_-) \;.  
\ee
 The coefficient $\beta$ is used to keep track of the logarithmic deformation; setting $\beta=0$ one recovers AdS$_3$. Hereafter, we take $\beta$ to be a small parameter since it can be regarded as a perturbation of the CFT by and irrelevant operator, see  \cite{Alishahiha2014} for a discussion. A curve in this space can be described by immersion functions
\be 
x^\mu (s) = \left(z(s),\,x_+(s),\,x_-(s)\right) \,,
\ee
where $s$ corresponds to curve's arc-length measured in units of $L$. Such a curve is determined by two independent functions $g(s)$ and $\psi(s)$, which we choose such that the tangent vector reads
\be
t^\mu=
\left(
\begin{array}{c}
-e^{f}\tanh\left( g+s\right) \\
\frac{1}{\sqrt{2}}e^{f+\psi}\sech \left( g+s\right)\\
-\frac{1}{\sqrt{2}}e^{f-\psi}\sech \left( g+s\right)\left(1+\frac{\beta}{2}e^{2\psi}f\right)
\end{array}
\right)\,,
\ee
where 
\be\label{f and g}
 f'(s)= - \tanh (g+s) \,.
\ee

We also need to find normal vectors such that 
\be \label{eta -}
n^{A\mu} n^{B}_\mu = \eta^{AB} = \left( \begin{array}{c c}
-1 & 0 \\
0 & 1
\end{array} \right) \,.
\ee
For instance, the normal vectors
\begin{align}\label{Logn}
n^{1\mu} &=
\left(
0,
\frac{1}{\sqrt{2}}e^{f+\psi},
\frac{1}{\sqrt{2}}e^{f-\psi} \left(1-\frac{1}{2}\beta e^{2\psi}f \right)
 \right)  \,, \\
n^{2\mu} &=
\left(
e^{f} \sech(\hat g),
\frac{1}{\sqrt{2}}e^{f+\psi}\tanh (\hat g),
-\frac{1}{\sqrt{2}}e^{f-\psi}\tanh (\hat g)\left(1+\frac{\beta}{2}e^{2\psi}f\right)
\right)  \nn\,,
\end{align}
where $\hat g=g+s$
fullfill this requirement.
The extrinsic curvatures associated to each normal direction can be elegantly written as
\be\label{logk}
k^A
=
\sech( \hat g)
\left( 
\begin{array}{c}
\dot\psi - \frac{\beta}{4 L} e^{2\psi} \tanh \left( \hat g\right) \\
\dot g- \frac{\beta}{4L} e^{2\psi}
\end{array}
\right) \,.
\ee
Notice that the frame \eqref{Logn} has a non-trivial curve torsion \eqref{tau d}
\be 
\tau=
- \dot\psi \tanh\left( \hat g\right) -\frac{\beta}{4 L} e^{2\psi}\sech^2\left( \hat g\right)\,.
\ee
The geodesic equations in the logarithmic background can be read from \eqref{logk}, and they are given by 
\begin{align}\label{logGeo}
\dot\psi =\frac{\beta}{4 L} e^{2\psi} \tanh\left( g+s\right)\,,\qquad \dot g = \frac{\beta}{4L} e^{2\psi}\,.
\end{align}
 Geodesics can then be found iteratively by expanding in $\beta$
\be\label{expa}
g(s)=\sum_{k=0} \beta^k\, g_k(s)\,,\qquad \psi(s)=\sum_{k=0} \beta^k\, \psi_k(s)\,.
\ee
For $\beta=0$ one recovers the AdS$_3$ geodesics which correspond to constant  $g_0$ and $\psi_0$. In particular, geodesics ending at unboosted intervals 
read $g_0=\psi_0=0$. 
The next order contributions to the latter are given by
\be\label{geo1 log}
g_1(s)=\frac{s }{4}\,, \qquad \psi_1(s) = \frac{1}{4} \log \cosh (s) \,.
\ee
In principle, one can continue this procedure to arbitrary order in $\beta$.

In what concerns the shape functionals, the crucial distinction between the logarithmic background and AdS$_3$ lies in 
the form of  the contractions of the ambient Riemann tensor, which read
\be
R^A_{\;\;A} = -\frac{4}{L^2} + \frac{\beta}{2 L^2} e^{2\psi} \sech^2(\hat g) \,,
\label{RAA}
\ee

\be
R^{AB}_{\;\;\;\;AB} 
=
  -\frac{2}{L^2} + \frac{\beta}{L^2} e^{2\psi} \sech^2 (\hat g) \,.
\ee
In constrast to AdS$_3$, these quantities are no longer constant and thus cannot be reabsorbed into $\lambda_0$. Indeed, the 
most general form of the functional~\eqref{eq:seff} in the logarithmic background~\eqref{log metric} is given by
\be
S[\Sigma]= \int ds \left( \hat{\lambda}_0 +\lambda_5' k_A k^A +\lambda_3'\frac{\beta}{2 L^2} e^{2\psi} 
\sech^2 \left(\hat g\right)
\right) \,,
\ee
where
\be
\hat\lambda_0=\lambda_0-\frac{2}{L^2}\left[3\lambda_2+2\lambda_3+\lambda_4\right] ~.
\ee
The shape equations \eqref{eq:eom1} can be written down explicitly and expanded in $\beta$ using Eq.~\eqref{expa}. Clearly, the zeroth order equation is of the form \eqref{Curve in MSS arc}, therefore it admits solutions like those discussed in Sec.~\ref{sec:curves} i.e. geodesics, constant mean curvature curves, wavelike or asymptotically geodesic. Above, we have seen that geodesics produce the right value of the central charge so we expand about these solutions. Expanding around the AdS$_3$ geodesic, at order $\beta$  we find  that the two shape equations decouple. The equation for $\psi_1(s)$ is solved automatically by the geodesic solution \eqref{geo1 log}, while for $g_1(s)$ we find
\begin{align}
0&=\hat{\lambda}_0 
\left(\dot g_1-\frac{1}{4}\right)
+
8\frac{\lambda_3 + 2 \lambda_4}{L^2}
\left(
 \dot g_1-\frac{1+\tanh^2 s}{8}
\right)
\nn\\
&+2
\frac{\lambda_5}{L^2}
\left(
\dddot g_1
-
2  \ddot g_1 \tanh s
+
2
(1+\tanh^2 s)
\left(\dot g_1-\frac{1}{4}\right)
\right) 
 \,.
\end{align}

The above equation is still rather complicated. Nevertheless, one can solve it analytically and express $\dot g_1$  in terms of hypergeometric and hyperbolic functions. We refrain from displaying the result here since it is not very illuminating. 
The key point is that for a suitable choice of integration constants
\be
 \dot g_1\to \frac{1}{4}\,,
\ee
 asymptotically ($s\to\pm \infty$).
Comparing with Eq.~\eqref{geo1 log} we see that this solution is asymptotically geodesic. Plugging it back into the functional we find that the only divergent contribution comes from the length term
\be
S[\Sigma] = 
2\hat{\lambda}_0 \ell_\Sigma
+ 
O(\beta^3) \,.
\ee 
Finally, we must relate $\ell_\Sigma$ to the UV cutoff $\epsilon$. To achieve this goal, we insert the asymptotically geodesic  
solution into
\be z(s)=Le^{f(s)}\,,
\ee
where $f(s)$ is given by Eq.~\eqref{f and g} and then invert $z(\ell_\Sigma) = \epsilon$.
This procedure, once more, should be performed iteratively in $\beta$. However, one finds that none of the subleading corrections contribute to the UV divergence, thus
\be
\ell_\Sigma
=   L \log \frac{\ell}{\epsilon} + O(\beta^3)\,,
\ee
where $\ell$ is the width of the interval in the boundary. To find this last result we computed also the second order corrections to the AdS$_3$ geodesic.
In the end, we are left with the remarkably simple result
\be\label{log EE}
S[\Sigma] = 
2L\hat{\lambda}_0 \log \frac{\ell}{\epsilon}
+ 
O(\beta^3) \,.
\ee 
As a matter of fact,  the above vanishes for the critical NMG couplings. The simplicity of Eq.~\eqref{log EE} stems from two interrelated reasons. First, the shape equations admit asymptotically geodesic solutions, this makes the contributions proportional to $k^A$ negligible in the UV. Secondly, the Riemann normal projections $R^{AB}_{\;\;\;\;AB} $ and $R^A_{\;\;A}$ when evaluated on the asymptotically geodesic solution approach a constant at the boundary. Thus, their contributions can be reabsorbed into the the definition of $\lambda_0$.

Notice that the universal contribution to the LogCFT entanglement entropy reported in equation (23)  of \cite{Alishahiha2014} does not match Eq.~\eqref{log EE}. We believe that the reason for this discrepancy is that the authors of
\cite{Alishahiha2014} overlooked the fact that their normal vectors don't satisfy Eq.~\eqref{eta -} and this omission pervades the rest of their computation. Notice that in~\cite{bianchini2014} the EE has an additional $\log(\log(\ell/\epsilon))$ divergence. This divergence can be traced back to logarithmic divergences in the two point functions of certain primary operators in LogCFTs, see \cite{Gurarie1993}. Apparently, these kind of divergences are not captured by the geometric formalism employed here. However, it is possible to link them to AdS$_3$-NMG at the chiral point by other means. In \cite{Grumiller:2009sn} these divergences where reproduced using the AdS/CFT recipe of quadratic fluctuations. It would be interesting to explore ways of incorporating that result into a geometric formalism. 

\section{Remarks on shapes in Euclidean space}
\label{sec:five}
It should not come as a surprise that the study of geometric functionals of the form~\eqref{eq:seff} and their associated shape equations~\eqref{eq:eom1} have some bearing on the investigation of classical problems of elasticity of surfaces and curves in $\mathbb{R}^3$. These kinds of questions are of interest in subjects ranging from the physics of polymers and membranes to pure differential geometry. The terms dependent on the ambient geometry's curvature drop out from~\eqref{eq:seff} leading to considerable simplifications; see~\eqref{eq:qwe}. Physical membranes can be modeled using smooth surfaces provided they display fluid-like behavior, which is realized by reparametrization invariance. Specifically, cell membranes can be described using the two dimensional fluid mosaic model proposed in~\cite{Singer720}. Based on this observation one can construct the functional that determines the shape of such membranes, which is the Canham--Helfrich~\cite{Helfrich1973, citeulike:7528087} free energy\footnote{For the moment we set the spontaneous curvature to zero. We shall discuss this quantity below.}
\be
S_\mathrm{\rm{CH}}[\Sigma] 
=  
\int_\Sigma d^2\sigma \sqrt{h}\,\left[\sigma+\frac{k_c}{4} (\tr K)^2+\bar{k}_c\det K \right]
\label{CanhamHelfrich}\,,
\ee
where $\sigma$ is the surface tension, while $k_c$  and $\bar{k}_c$ are known as the bending rigidities. Notice that the third term in the above functional, called the Gaussian curvature, is a total derivative. Nevertheless, using the relation
\be 
\det K =\frac{1}{2}\left[(\tr K)^2-(\tr K^2)\right]\,,
\ee
it is straightforward to relate the surface tension and the bending rigidities to the $\lambda_i$ coefficients in \eqref{eq:seff}. Interestingly, a special case of~\eqref{CanhamHelfrich} yields the only conformally invariant combination of quadratic invariants, namely, the Willmore energy
\be
S_\mathrm{W}[\Sigma] 
=  
\int_\Sigma d^2\sigma \sqrt{h}\,\left[\frac{1}{4}(\tr K)^2-\det K\right]
\label{Willmore}\,.
\ee
The shape equation corresponding to this functional can be obtained from~\eqref{eq:eom1}, and it reads
\be
\Delta\tr K-\frac{1}{2}(\tr K)^3+\tr K \tr(K^2)=0\,.
\ee
The study of the solutions of this equation, called Willmore surfaces, has been the subject of recent and groundbreaking studies in mathematics~\cite{Marques2013}.

Another interesting problem is the study of curves in $\mathbb{R}^3$, where the action is essentially~\eqref{mss curve}
\be\label{eq:EB}
S_\mathrm{EB}[\Sigma] 
=
\int_\Sigma d\sigma\sqrt{h}\left[\lambda_1+\lambda_2
\tr k_A \tr k^A \right]\,.
\ee
Finding extrema of this functional is a problem with a longstanding tradition. In fact, for fixed total length, this variational problem was proposed by Daniel Bernoulli to Leonhard Euler in 1744. (See~\cite{Singer2008} and references therein.) Physically speaking, $\lambda_1$ encodes the line tension while $\lambda_2$ is the corresponding one dimensional bending rigidity, which quantifies the resistance to bending posed by an infinitesimal cross section of the material.

There is one aspect in which the study of curves in $\mathbb{R}^3$ is richer than that of surfaces. Indeed, since the relevant codimension is $d-p=2$ there is a non-trivial gauge freedom in the choice of normal frames. (See Section~\ref{sec: gauge invariance}.) In this case the normal gauge group corresponds to $O(2)\simeq U(1)$. In fact, this is the simplest case where a non-vanishing extrinsic torsion can arise, leading to the introduction of the curve torsion~\eqref{tau d}.
There is a particular frame, or gauge choice, that plays a central role in the theory of curves, the Frenet--Serret (FS) frame~\cite{DoCarmo1976}. Once we have chosen an arclength parametrization, the Frenet--Serret frame is engineered in such a way that the total extrinsic curvature is captured by a single normal direction. Often the price to pay for this choice is to have a non-vanishing extrinsic torsion. On the other hand, for closed curves it is always possible to find a normal frame where $\tau=0$. In this frame, the geometry of the embedding is entirely described by the two extrinsic curvatures $\tr k_1$ and $\tr k_2$, which are in general non-vanishing. Hence, we must make a compromise, either a single extrinsic curvature and torsion or two extrinsic curvatures and vanishing torsion. Evidently, these two options are connected by a gauge transformation. Indeed,
\bea 
k_\mathrm{FS}^2 &=& k_1^2 + k_2^2  \,, \label{eq:kfs}
\\
\tau_\mathrm{FS} &=& \frac{k_1 \dot k_2-k_2 \dot k_1}{k_1^2+k_2^2}\,.  \label{eq:tfs}
\eea
Recall that we are in the arclength parametrization, hence, the absence of traces in the above expression. Notice that whenever a portion of 
the curve is planar one of the $k$s is zero and  hence $\tau_\mathrm{FS}=0$. One must be careful though in the case of straight lines where both $k$s vanish and the Frenet--Serret frame is ill-defined. This construction can be extended to embeddings where $\mathbb{R}^3$ is replaced by a general smooth three dimensional manifold~\cite{Singer2007}. As a matter of fact, we can follow this reasoning whenever $d-p=2$.

Observe that  \eqref{eq:kfs} is a gauge invariant quantity, being simply the low dimensional analogue of $\tr K_A \tr K^A$. Meanwhile, it ought to be clear that \eqref{eq:tfs} is not gauge invariant. However, we can incorporate $\tau_\mathrm{FS}$ into a gauge invariant combination by considering an invariant term of the form
\be
 h^{ij}(\tilde{D}_i^{BA} \tr K_A)(  \tilde{D}_j^{CD} \tr K_D) \eta_{BC}\,.
\ee
Which in the present setup reduces to 
\be
\dot k_\mathrm{FS}^2 + \tau_\mathrm{FS}^2 k_\mathrm{FS}^2 
=
\dot k_1^2
+
\dot k_2^2  \,.
\label{dkfs}
\ee
Analogous expressions where found in \cite{Arreaga:2001ma}. Equation \eqref{dkfs} is the simplest and most direct application of the gauge invariance principle discussed in Section~\ref{sec: gauge invariance}. The upshot is that an action functional can't depend arbitrarily on the torsion without leading to a breakdown of gauge invariance. At this point, this remark might seem trite. However, it is a rather important fact and there is a large body of literature that doesn't seem to do justice to it. 


It is often the case in physics that effective descriptions must take into account possible explicit symmetry breaking terms which can be explained only by considerations originating at smaller scales. While reparametrization invariance on $\Sigma$ is a necessary symmetry of any geometrical problem, this is not the case for the normal bundle's gauge invariance. In fact, in the two systems discussed above it is possible to incorporate physically sensible terms that break gauge invariance. For example, in its  original formulation the Canham--Helfrich free energy contains an extra term $\tr K_0$ 
\be
S_\mathrm{\rm{CH}}[\Sigma] 
=  
\int_\Sigma d^2\sigma \sqrt{h}\,\left[\sigma+\frac{k_c}{4} (\tr K- \tr K_0)^2+\bar{k}_c\det K \right]
\label{CanhamHelfrich1}\,,
\ee  
which is a constant called the \emph{spontaneous curvature}. To understand the meaning of this quantity, consider the problem of finding closed surfaces of fixed area which extremize the functional~\eqref{CanhamHelfrich1}. Clearly the answer to this question recalls the constant mean curvature solutions wherein the mean curvature matches the spontaneous curvature ($\tr K = \tr K_0\,$). 

Similarly, the Euler--Bernoulli model \eqref{eq:EB} can be modified to support non-gauge invariant contributions. For example, imagine that the curve is an effective description of a developable, infinitely thin ribbon. Ribbons, however thin, are two dimensional objects, due to this fact they inherit a preferred frame onto the one dimensional description. The normal vector to the ribbon becomes one of the vectors of the normal frame, thus fixing up to a residual $\mathbb{Z}_2$ a natural frame in the normal bundle, which is customarily referred to as the \emph{material frame}. The existence of a preferred frame is in flagrant violation of gauge invariance but clearly the physics justifies its existence. Now, the only term quadratic in the curvature of the two dimensional action reduces to the one dimensional Sadowsky--W\"underlich~\cite{sadowsky1930theorie,Wunderlich1962, fosdick2015mechanics} functional, which in the Frenet--Serret frame reads
\be\label{Sad Wun}
S_2 [\Sigma] 
\simeq
\int_\Sigma ds
\frac{(k_\mathrm{FS}^2+\tau_\mathrm{FS}^2)^2}{k_\mathrm{FS}^2} \,.
\ee
Interestingly, when~\eqref{twisting} is evaluated in the material frame, it expresses the number of times a physical ribbon or wire winds onto itself.
The message we wish to convey with these examples is that if a geometric action must break gauge invariance it has to do so for a physical reason. Once the requirement of gauge invariance is forsaken the landscape of allowed terms in any effective action grows significantly and physical intuition becomes the only guiding principle.

\begin{figure}[h]
\centering
\includegraphics[width=1\textwidth]{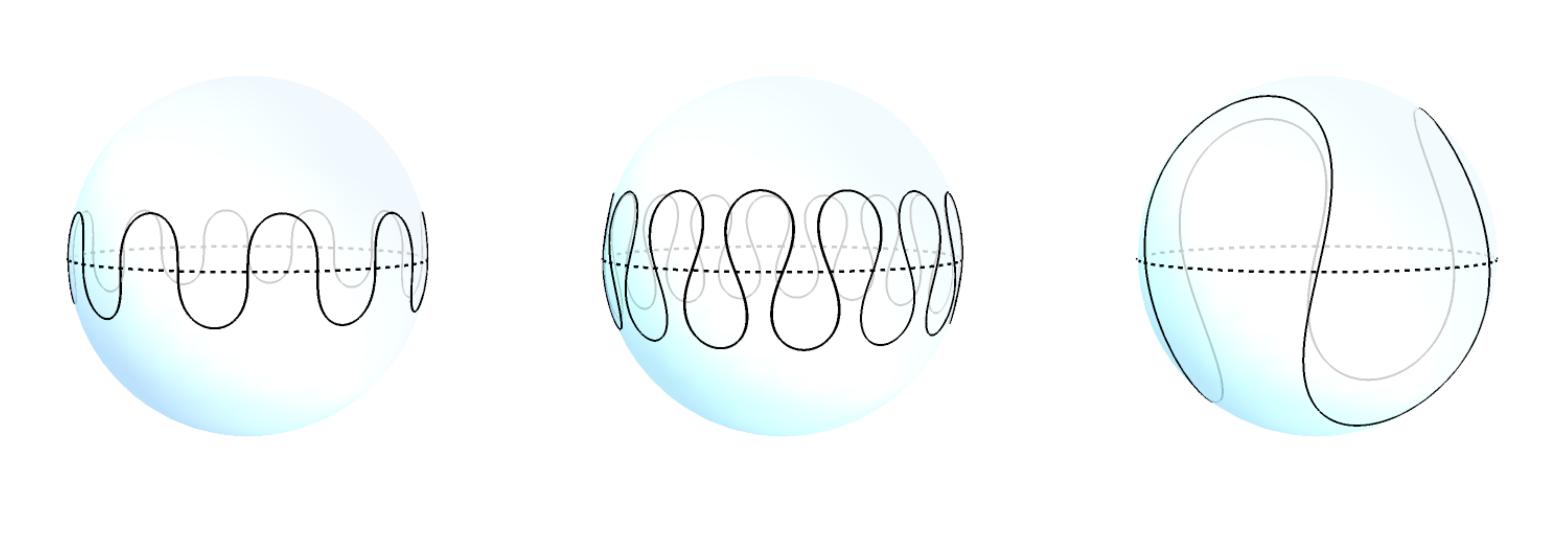}
\caption{Elastica on a sphere.}
\label{fig:wavysph}
\end{figure}

Finally, let us touch upon another interesting class of examples, namely, curves embedded into surfaces. We can take two perspectives when handling these problems. For instance, we could treat the problem \emph{intrinsically}, \textit{i.e.}, by viewing the curve in question as $\Sigma$ and the surface as the ambient manifold. Alternatively, we could regard it as the study of a curve in $\mathbb{R}^3$ where gauge invariance is broken by selecting the normal vector of the surfaces as one of the members of the normal frame \cite{Guven2012}. Finding these \emph{doubly embedded} elastica is rather non-trivial. Even if the surface is symmetric enough to allow for an analytic expression of the extrinsic curvatures (see \textit{e.g.}, Section~\ref{sec:curves}), constructing the actual curves is quite involved but can be done analytically. Indeed, by a procedure parallel to the one outlined in \ref{sec:trK}, one can reproduce the elastica on a sphere found in \cite{Langer1984}; see Figure~\ref{fig:wavysph}. It is natural to wish to explore this further. The geometric formalism we have discussed can be easily adapted for the study of more complicated settings. One could, for instance, study the coupled shape equations on mobile surfaces, in fact this problem finds applications in the theory of membranes \cite{Giomirspa20110627}. Also, it is possible to explore the shape equations for a curve on a time-dependent surface, see \cite{2016JNS...tmp...35Y} for work in this direction. We hope that some of the lessons discovered working in the latter setup will have some relevance in the study of entanglement entropy in out of equilibrium systems via holography, see \cite{2007JHEP...07..062H, 2015arXiv151205666C}.

\section{Summary and discussion}\label{sec:six}

This work is devoted to the study of geometric functionals and their extrema. More concretely, we address the question of which shape a manifold is compelled to take if it extremizes a given geometric functional. Our investigations are driven by physical interests but take a purely geometric approach. The use of a geometric perspective has a twofold benefit: it gives results of wide generality and yields equations with a meaningful structure. We find that the geometries that extremize functionals of the form \eqref{eq:seff} obey the \emph{shape equations} \eqref{eq:eom1}, which depend solely on well-defined geometrical objects. In order to deduce and solve these equations, a fair deal of geometric technology is needed; we have placed the necessary concepts in \ref{sec:app geometry}. An interesting offshoot of these geometrical disquisitions is the realization of the existence of an underlying gauge freedom in the choice of normal directions. We discuss this in Section \ref{sec: gauge invariance}. This gauge freedom implies the existence of a connection, which interestingly corresponds to the extrinsic torsion \eqref{eq:tor}. Once the torsion is viewed as a connection a number of questions in holography and elastica theory become more transparent.  With the exception of curves, in general, it is not possible to set the torsion to zero via gauge choice. Hence, we wish to stress the importance of not overlooking the existence of this quantity. In fact, one expects the shape equations to be fully covariant under gauge transformations and indeed this is the case. Even though many of the tools used in deriving the shape equations were previously derived in \cite{PhysRevD.48.4604, PhysRevD.51.6736, Capovilla:2006te}, we present our independent derivations in considerable detail in \ref{eom}. We believe that the contents of Section \ref{sec: gauge invariance}, \ref{sec:app geometry} and \ref{eom} provide a useful summary for someone wishing to enter this field.

In determining the shapes of extrema, the geometrical character of the equations~\eqref{eq:eom1} is extremely advantageous. If we were to write these, as equations for the shape itself, clearly the result would be a rather complicated system of fourth order, non-linear differential equations. In fact, in a number of works, \emph{e.g.}, \cite{Erdmenger2014, Hosseini2015, Ghodsi:2015gna}, the path taken was the following: first one chooses a parametrization for the submanifold in question, then one computes the geometric quantities appearing in the action, and finally, one derives the Euler--Lagrange equations of motion for the functions that define the parametrization. Finding all the solutions to the resulting system, even in simple scenarios, seems a daunting, if not unsurmountable, endeavor. However, in some cases, using the geometric form of the shape equations one can separate this question into a hierarchy of tractable problems. For instance, if the ambient manifold is maximally symmetric then the shape equations reduce to \eqref{shape mss}, which is a system of second order equations for the extrinsic curvatures. Right away, it is possible to draw interesting conclusions from these equations, such as the conditions needed for a minimal submanifold to be an extremum; see~\eqref{cond minim}. If one manages to find the extrinsic curvatures, then computing the actual shape of the submanifold reduces to another second order problem. Following seminal work by Langer and Singer \cite{Langer1984}, in Section \ref{sec:curves} we show how to calculate analytically the extrinsic curvature of a curve in a maximally symmetric surface. Then, in  \ref{sec:trK} we carry out the second step explicitly by inverting the extrinsic curvature in the case of Lobachevsky space $\mathbb{H}^2$. The final result of this procedure appears in~\eqref{wavy explicit} and is depicted in Figure~\ref{fig:wavy}. We want to stress that, in this context, the procedure outlined above allows one to find all the solutions to the shape equations analytically.

Afterwards, we apply the above formalism to specific physical setups. First, we consider the problem of computing entanglement entropy from a holographic perspective. The functional that computes the entanglement entropy for quantum field theories whose holographic dual is a gravity theory of the form \eqref{HC} is given by \eqref{Dong}. Clearly, this functional is a particular case of \eqref{eq:seff} and all the general results concerning the shape equations are applicable to its extrema. Moreover, in \cite{Dong2014}  it was  shown the shape equations corresponding to \eqref{HC} match the equations proposed in \cite{Lewkowycz:2013nqa}, which are known to be satisfied by the right entangling surface. Thus, we learn that to obtain the entanglement entropy we must evaluate the functional \eqref{eq:seff} on one of its extrema.  The question of which of the potentially infinite possible extrema yields the correct value of the entanglement entropy remains to be settled. In analogy with the Ryu--Takayanagi prescription one would expect the right surface to be a minimum of the functional.

In Section \ref{EE 3d} we address the question of minimality in the context of four derivative gravity in three dimensions, where we can apply the findings of Section \ref{sec:curves} straightforwardly. If we were to compute the entanglement entropy for an interval in the boundary CFT, thanks to the results in Section~\ref{sec:three} we can construct all the possible static entangling curves in AdS$_3$. See Figure~\ref{fig:u} for interesting examples. The simplest types of entangling curves are those with non-zero constant mean curvature and geodesics. In the context of New Massive Gravity, it was argued in \cite{Ghodsi:2015gna} that while the geodesics yield the correct value for the entanglement entropy, they cannot be global minima since their on-shell value is larger than that of curves with non-vanishing constant mean curvature. Here, after showing that geodesics provide the right value for the entanglement entropy for any four derivative theory in three dimensions, we evaluate \eqref{eq:seff} on all of its extrema. We discover that in New Massive Gravity, the functional always takes its largest value on the geodesics. This is not what we naively expect based on the Ryu--Takayanagi prescription. To our knowledge, this is the first case where all the possible entangling curves are known for a higher curvature theory. Having analytic control over all these curves opens some interesting avenues to explore. For instance, it could help in the search for a prescription to find the correct entangling surface in more general settings. Also, one could investigate whether these new entangling curves have interesting information theoretic interpretations along the lines of \cite{Czech:2014tva}. Moreover, we might be able to understand analytically non-geodesic curves in Topological Massive Gravity as those studied in \cite{Castro:2014tta}.

The geometric formalism discussed in this work can be applied naturally to problems concerning elastica and membranes. These are questions regarding surfaces and curves immersed in Euclidean space. This formalism provides the tools to clarify certain aspects that are sometimes, to our view, overlooked in the literature. A crucial point is the explicit appearance of torsion in energy functionals such as the Sadowsky--W\"underlich energy \eqref{Sad Wun} for a curve. As discussed in Section \ref{sec: gauge invariance}, the extrinsic torsion transform as a gauge field under rotations of the normal frame. Therefore, a functional such as \eqref{Sad Wun} isn't invariant under choices of normal frame. This implies the existence of preferred frames and this must be justified. Indeed, in the Sadowsky--W\"underlich formalism we treat ribbons ($p=2$) as curves ($p=1$) and the presence  of a preferred frame is inherited from the higher dimensional origin of the problem. The stance we take is that gauge invariance should be used as a guiding principle to construct effective actions and the addition of terms breaking it must be advanced on physical grounds.

\subsection{Future directions}

Above, we pointed out some possible applications of the shape equation formalism beyond the scope of this work. Now, we list other potential directions to explore.
\begin{itemize}
\item We showed that for AdS$_3$ geodesics are the right entangling curves. However, we have seen that minimal submanifolds \eqref{trK0} aren't always extrema even for maximally symmetric spaces. Moreover, for generic ambient manifolds there is no guarantee that even geodesics are extrema. Thus, we might wonder which criterion must be used to select the right entangling curve if geodesics aren't extrema. This problem was partially addressed in \cite{Hosseini2015} for the case of a hairy black hole in New Massive Gravity \cite{Oliva:2009ip} for which geodesics don't satisfy the shape equations. We hope that with the analytic understanding developed here, this question can be tackled in a more systematic manner. Moreover, we would like to explore other scenarios where this issue is present such as Lifshitz \cite{Hosseini:2015gua, Basanisi:2016hsh} and logarithmic metrics \cite{Alishahiha2014}.

\item The formalism discussed in the present work is valid for arbitrary dimension and codimension. Therefore, it is natural to go on and investigate higher dimensional settings. There are two possibilities that come to mind right away. First, recall that the crucial point leading to the analytic expression for extrema such as \eqref{wavy explicit} was the hierarchical splitting of the shape equations, namely, the fact that from the shape equations one can find the extrinsic curvatures first and then from these find the shape of the submanifold. From \eqref{shape mss} we see that this splitting occurs for any maximally symmetric ambient space. An interesting feature of this equation is that for $p\geq 3$ minimal submanifolds are not necessarily extrema, unless condition \eqref{cond minim} is satisfied. As we have seen, for curves this equation can be integrated in terms of elliptic functions. Of course, one wonders whether similar progress can be done in higher dimensional theories. The other possibility comes from considerations regarding the Killing vectors. The existence of Killing directions in the ambient manifold can lead to trivializations of the normal and tangent bundles.  This might lead 
to a dimensional reduction of the problem. In fact, we have used this implicitly in Section \ref{EE 3d} where we reduced a problem in AdS$_3$ to one in Lobachevsky space. (This is explained at the end of  \ref{eom}.) We believe that this feature of dimensional reduction also deserves further attention. Moreover, as seen in \ref{sec:trK}, the existence of Killing fields was crucial in inverting the extrinsic curvature.

\item There are certain questions that might require numerical techniques but appear to be rather compelling. For example, we could consider the shape equations for a submanifold immersed in a time dependent ambient geometry. Stimulating work in this direction can be found in \cite{2016JNS...tmp...35Y}. Moreover, it would be interesting to apply our general geometric considerations in building action functionals where objects of different dimensionalities interact. In particular, the construction of a configuration energy of a two-component elastic membrane with non trivial one dimensional interface bending rigidity is an open interesting problem.

\item Furthermore, it would be interesting to study the behavior of the shape equations, and the generalized curvature identities, under conformal maps. These transformations can be used to build bridges between different geometrical problems. Then, these connections can be used to carry insights from one problem to the other. This is the case, for example, for the question of finding minimal surfaces in $\mathbb{H}^3$ and that of computing Willmore surfaces in $\mathbb{R}^3$. As shown in \cite{Alexakis2010}, these problems transform into each other under conformal maps, this observation has been applied in the context of holography in \cite{Fonda:2015nma}.

\item
We view extrema (\textit{i.e.}, the solutions to the shape equations) as fixed points of geometric flows.
Whereas mean curvature flows perform a steepest descent on the area, we can use steepest descent to extremize other geometric quantities.
In particular, just as minimal surfaces are fixed points of mean curvature flows, Willmore surfaces are fixed points of Willmore flows, etc.
Recasting constrained optimization problems in terms of geometric flows has several natural advantages.
It is ideal, for example, for realizing numerical solutions.
No matter the surface from which one starts, the flow (if it is convergent) will eventually lead to the desired extrema.

The concept of geometric flows is very interesting per se and is a rich vein that has been much tapped in various mathematical contexts. 
We may consider \textit{intrinsic} geometric flows, like Ricci flow, where the rate of change of the metric tensor at a given point on a manifold is proportional to the Ricci tensor:
\be
\frac{dg_{\mu\nu}}{d\lambda} = -2R_{\mu\nu}(g) ~,
\ee
where $\lambda$ is some parameter along the flow.
Fixed points of this flow are necessarily Ricci flat geometries.
If we imagine the manifold as embedded in a larger one, this flow is essentially a modification of~\eqref{hnLie}, where one replaces the extrinsic curvature with the intrinsic Ricci tensor and has to imagine a normal displacement as a shift in the parameter $\lambda$.

Perelman's solution to the Poincar\'e conjecture proposes an entropy functional
\be
{\cal F} = \int_M dV\ e^{-f}\ (R+(\nabla f)^2) ~,
\ee
which is dilaton gravity on a Riemannian manifold, and considers gradient flow equations associated to variations of this entropy~\cite{Perelman2002}.
The extrema that are the endpoints of the flow will in general not be minimal surfaces or Ricci flat geometries.
It would indeed be enlightening to understand this in the context of this paper.
Moreover, it is very interesting to contemplate flows that mix the purely intrinsic Ricci flow with extrinsic flows such as those we have been discussing. 

Ricci flow \`a la Perelman is essentially the same as the renormalization group evolution of a non-linear sigma model on a string worldsheet with target space metric $g_{\mu\nu}$~\cite{Friedan:1980jf}.
The connection between optimization problems couched in the language of gradient flows and the renormalization group has not been fully explored within string theory or in terms of the gauge/gravity correspondence.
Initial efforts in this directions appear in~\cite{Jackson:2013eqa}.
We have noted that there is a gauge redundancy in the description of the system; this should ultimately be related to diffeomorphism invariance in the bulk and scheme independence in the dual CFT~\cite{Banks:1987qs,Hughes:1988bw,Balasubramanian:2000pq}.
\end{itemize}

Certainly, there are a plethora of interesting questions in this subject that deserve to be addressed. In the present work we hope to have provided a clear picture of the basic ingredients needed to treat questions regarding the shape of things. We would like to finish by saying that, pedestrians that we are, we are joyful to have caught glimpses into to the beautiful landscape of geometry and we hope to have conveyed some of this experience to our readers.

\section*{Acknowledgements}

The work of PF was primarily supported by the Angelo Della Riccia Foundation for the duration of this project. PF was partially supported by The Netherlands Organization for Scientific Research (NWO/OCW). Moreover, PF thanks the University of the Witwatersrand for support and hospitality during the initial phase of this project. 
VJ is supported by the National Research Foundation and the South African Research Chairs Initiative.
VJ thanks the academic staff and the string group at Queen Mary, University of London for its always generous hospitality.
The work of AVO is based upon research supported in part by the South African Research Chairs Initiative of the Department of Science and Technology and National Research Foundation. AVO's research is also supported by the NCN grant 2012/06/A/ST2/00396. AVO wishes to thank the theory group at CERN and the mathematics department at IST Lisbon for their hospitality during the development of this work. Also, AVO is grateful to Caravasar where he carried out some of the final stages of this project. 
We wish to acknowledge
Mohsen Alishahiha,
Jay Armas,
 Luca Giomi,
Shajid Haque,
Mohammad Reza Mohammadi Mozaffar,
Gon\c{c}alo Oliveira, and
Flavio Porri
for enlightening conversations and correspondence, as well as for helpful comments on earlier versions of this manuscript. 
PF thanks Marco Raveri for very interesting discussions halfway through the completion of this work.


 \appendix

\section{Geometric technology}\label{sec:app geometry}

\begin{figure}[h]
\centering
\includegraphics[width=.65\textwidth]{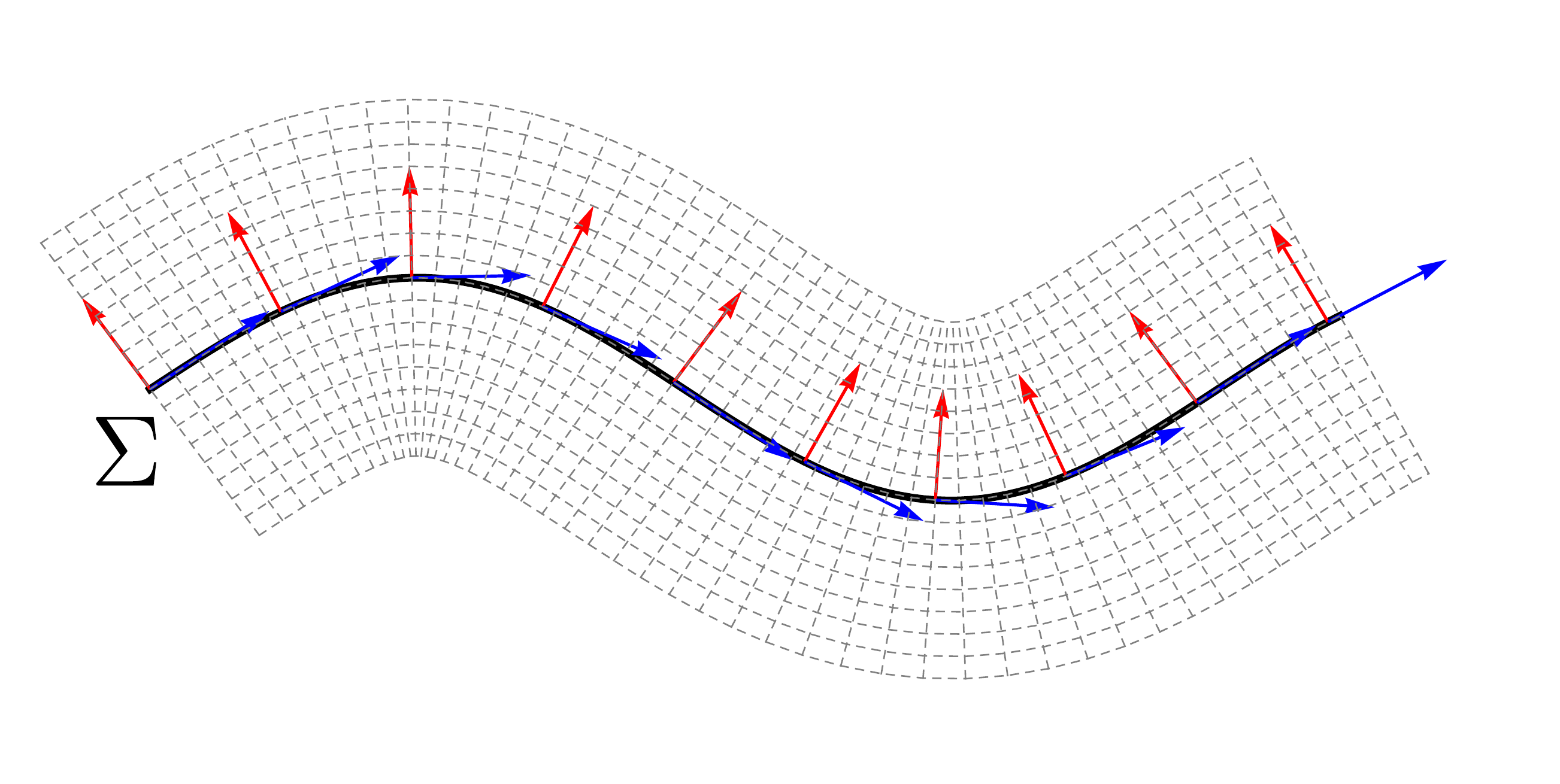}
\caption{Graphic example of the coordinate frame used in this section. Here $\Sigma$ is a curve embedded in the two dimensional Euclidean plane, for which we depict in blue and red respectively the tangent and normal vector fields. The dashed gray lines show the coordinates grid. At each point outside of $\Sigma$, $TM$ clearly decomposes in a normal and tangent direction.}
\label{fig0a}
\end{figure}

In this Appendix we explore some of the geometrical properties of the setup described in Section~\ref{sec:two}.
To start, let us define a suitable coordinate system in the neighborhood of $\Sigma$.
The relevant coordinates are constructed as follows.
Consider the family of integral curves generated by the span of $n^A$ emanating from $\Sigma$, see Figure~\ref{fig0a}.
If $y\in M$ is a point in the neighborhood of $\Sigma$, then it lies in one and only one of the aforementioned integral curves; call this $\gamma_y$. 
The coordinates we shall use to label $y$ are those of the point where $\gamma_y$ meets $\Sigma$ together with the \emph{distances} in each of the directions $n^A$ which $\gamma_y$ had to traverse to reach $y$.
Infinitesimally,  we can write $y$ as 
\be\label{defining y}
y^\mu = x^\mu(\sigma^i)+\varepsilon^A n_A^\mu\,,
\ee
and thus, we assign to $y$ the coordinates $\{\sigma^i,\varepsilon^A\}$.
The tangent bundle of $M$ restricted to $\Sigma$ can be decomposed naturally into tangent and and normal directions using the basis  $\{t_i^\mu,n^\mu_A\}$.
Now, the integral curves $\gamma_y$ can be regarded as maps in $M$ taking $x\mapsto y$ and can be used to extend the vector fields $t_i^\mu$ and $n^\mu_A$ away from $\Sigma$ via push forward. With this construction, we extend the normal/tangent factorization of the tangent bundle on $\Sigma$ to a neighbourhood of  $\Sigma$. Furthermore, this construction can be used to extend $h_{ij}$ and $K^A_{ij}$ away from $\Sigma$ in the same neighbourhood. From another point of view, notice that the neighbourhood in question is naturally foliated by push-forwarded copies of $\Sigma$. The extended $h_{ij}$ and $K^A_{ij}$ correspond to the induced metrics and extrinsic curvatures of the leaves of this foliation. It must be pointed out that this construction depends on $n_A^\mu$ hence, it is associated with a gauge choice. This will manifest itself in the form of non-gauge covariant intermediate results. However, just as in familiar gauge theory computations, physical ( in the present case geometrical) quantities must transform covariantly. Note that some of the intermediate results in the following could be in principle made covariant by incorporating some of the connection terms into the definition of the normal variations, as it was done in \cite{PhysRevD.51.6736}. Abusing slightly the notation, we denote the extended fields by $t_i^\mu$, $n^\mu_A$, $h_{ij}$ and $K^A_{ij}$ as well.

By construction, the Lie derivatives obey
\be
\mathcal{L}_{A} t^\nu_i \equiv\mathcal{L}_{n^\mu_A} t^\nu_i  = 0 \qquad  \mathcal{L}_{t^\nu_i} n^\mu_A \equiv  \mathcal{L}_{i} n^\mu_A = 0 \,.
\label{lien}
\ee
Moreover, since $\left[ t^\mu_i, t^\nu_j\right]=0$ on $\Sigma$ and push forwards commute with Lie brackets then 
\be
\mathcal{L}_{i} t^\nu_j = 0 \,,
\label{liet}
\ee
holds in the entire neighborhood.
Since $M$ is a Riemannian manifold, equipped with a metric $g_{\mu\nu}$ and a torsion-free Levi-Civita connection, we can compute Lie derivatives using covariant differentiation on $M$.
In particular we can rewrite~\eqref{lien} using the connection on $M$, from which we then deduce
\be \label{conm}
t^\nu_i \nabla_\nu n^{\mu A} = n^{\nu A}\nabla_\nu t^\mu_i  \,.
\ee
We see that the rate of covariant change of a normal vector along a tangential direction is the same as the rate of change of a tangential vector along a normal direction.
Notice that in general 
\be
\mathcal{L}_{A} n_B^\nu \neq 0 \,,
\label{lienn}
\ee
we will study the explicit form of this expression in some detail below. 

Subsequently, we compute the normal and tangent derivatives of vector fields using the above construction.
As we shall see, all the derivatives can be expressed in terms of the following four objects:
\begin{itemize}
\item \textit{Intrinsic connection}
\be
\tilde{\Gamma}^k_{ij} = t^{\mu}_i t^{k\nu} \nabla_\mu  t_{j\nu} \,.
\ee
\item \textit{Extrinsic curvature}
\be\label{Extrinsic curvature}
K_{ij}^A = t^{\mu}_i t_j^\nu \nabla_\mu   n^{A}_\nu \,.
\ee
\item \textit{Extrinsic torsion}
\be
T_i^{AB} = n^{A\mu} t_i^\nu  \nabla_\mu   n^{B}_\nu \,.
\ee
\item \textit{Normal connection}
\be
\Theta^{AB}_{\;\;\;\;C} = n^{A\mu} n_C^\nu \nabla_\mu   n^{B}_\nu \,.
\ee
\end{itemize}
In terms of these quantities, we can write 
\be  
\mathcal{L}_{A} n^{B\nu} = T_i^{[AB]} t^{i\nu} + \Theta^{[AB]}_{\;\;\;\;C} n^{C \nu} \,.
\label{liennn}
\ee

Finally, using~\eqref{lien},~\eqref{liet} and~\eqref{lienn} and the above definitions the tangential and normal derivatives read
\begin{table}[h]
\centering
\begin{tabular}[c]{r c c c l}
&&&&\\
$t^{\mu}_i \nabla_\mu  t_j^\nu
$&=&$ \tilde{\Gamma}^k_{ij} t^{\nu}_k $&$-$&$ K_{ij}^A n_A^\nu $ \\ 
$n^{A\mu} \nabla_\mu   t_i^\nu
$&=&$  K_{ij}^A t^{j \nu} $&$-$&$ T_{i}^{AB} n_B^\nu $ \\ 
$t^{\mu}_i \nabla_\mu   n^{A\nu}
$&=&$  K_{ij}^A t^{j \nu} $&$-$&$ T_{i}^{AB} n_B^\nu $ \\ 
$n^{A\mu} \nabla_\mu   n^{B\nu}
$&=&$ T_{i}^{AB}  t^{i\nu} $&$+$&$ \Theta^{AB}_{\;\;\;\;C} n^{C\nu} $ \\&&&&\\  \end{tabular}
\caption{Summary table of the tangent and normal decompositions of the four connection forms.\com{(we never called them elsewhere like that, so maybe we should use another name, although geometrically correct)}
From $\mathcal{L}_{i} t^\nu_j=0$ it follows that $K_{ij}^A$ and $\tilde{\Gamma}^k_{ij}$ are both symmetric in lower indexes.
From $n^{A\mu} t_{i\mu}=0$ it follows that the same coefficient $K_{ij}^A$ shows in the first and second line with opposite sign, and similarly the same coefficient $T_i^{AB}$ shows in the third and the fourth line.
From $\mathcal{L}_{i} n^{A\nu}=- \mathcal{L}_{A} t^\nu_j=0$ it follows that the second and third lines are equal.
From $\nabla_\mu \eta^{AB}=0$ it follows that $T_i^{AB}=-T_i^{BA}$ and $\Theta_{ABC}=-\Theta_{ACB}$.}\label{tablecf}
\end{table}

\subsection{Curvature identities}\label{curvature identities}
Our next task is to derive a number of consistency conditions on the curvatures which any embedding ought to satisfy.\footnote{
The reader is invited to keep the notation of the table in Section \ref{notation} in mind.}
These conditions can be found by computing the appropriate Lie derivatives, employing the identities~\eqref{lien},~\eqref{liet},~\eqref{liennn} and applying the Leibniz rule judiciously.
As a first step, we calculate the tangential and normal Lie derivatives of the induced metric and find the relations
\bea
\mathcal{L}_k h_{ij} &=& \tilde{\Gamma}_{ik}^l h_{lj} +\tilde{\Gamma}_{jk}^l h_{il} \,,
\label{htLie}\\
\mathcal{L}_A h_{ij} &=& 2 K_{ij}^A\,,
\label{hnLie}
\eea
where we used $\mathcal{L}_v g_{\mu\nu}= \nabla_\mu v_\nu + \nabla_\nu v_\mu$.
Notice that the first of the above identities captures the compatibility of $\tilde\nabla$ and $h_{ij}$.
The next step is to compute the tangential and normal Lie derivatives of the connection forms.
The results are displayed below:
the tangential derivatives are summarized in Table~\ref{tabletangent} while the normal ones are in Table~\ref{tablenormal}.

 With respect to the metric of the ambient space $M$, the extrinsic curvatures, extrinsic torsions and induced connections are scalar objects.
 Since for such objects $\mathcal{L}_i F = t_i^\mu \nabla_\mu F = \partial_i F$ holds, by simply antisymmetrizing~\eqref{Liei1} in $i \leftrightarrow l$ we find the Gauss relation
\be
\mathcal{R}_{j k il}
=
R_{j k il}
+
K_{[i j}^A K_{k l]}^B  \eta_{AB} \,.
\label{GClieOLD}
\ee
From which the contracted identities
\bea 
&&\mathcal{R}_{ij}
=
R_{ij}
-
R_{i\;\;\;j A}^{\;\;A} 
+
K_{ij}^A \tr K_A  
-
K_{ik}^A K^B_{jl} \eta_{AB} h^{kl} \,,
\label{GClie1}
\\
&&\mathcal{R} 
=
R
-
2 R_A^{\;\;A}
+
R_{AB}^{\;\;\;\;\;AB}
+
\tr K_A \tr K^A
-
\tr (K_A K^A)
\,,
\label{GClie2}
\eea
follow.
In turn, the Codazzi--Mainardi equation
\be
R^A_{\;\;j ik }
=
\tilde{\nabla}_{[k} K_{i]j}^A
-
K_{[kj}^B T_{i]}^{AC} \eta_{BC} \,,
\label{CMlieOLD}
\ee
can be found from~\eqref{Liei2} by a $i \leftrightarrow k$ anti-symmetrization.
Meanwhile, from~\eqref{Liei3} we obtain a generalized version of the Ricci equation
\be 
R^{AB}_{\;\;\;\;\;ij}=\tilde{\nabla}_{[i} T_{j]}^{AB}-K_{[ik}^A K_{j]l}^B h^{kl}
-
T_{[i}^{AC} T_{j]}^{BD} \eta_{CD} \,.
\label{RclieOLD}
\ee 

\begin{table}[h]
\centering
\begin{align}
&\mathcal{L}_l \tilde{\Gamma}_{ij}^k = t^{\mu}_i t^{k \nu} t^{\rho}_l \nabla_\rho \nabla_\mu t_{j \nu}+\tilde{\Gamma}^m_{i[l}\tilde{\Gamma}^k_{mj]}-K^A_{i[l} K^B_{m j]} \eta_{AB} h^{km}\label{Liei1} ~,\\
&\mathcal{L}_k K_{ij}^A =
t^{\mu}_i t^{\nu}_j t^{\rho}_k \nabla_\rho \nabla_\mu n^A_{\nu}
+
\tilde{\Gamma}^{m}_{(ik} K_{mj)}^A
+
K^B_{(ik} T_{j)}^{AC} \eta_{BC}
\label{Liei2} ~, \\
&\mathcal{L}_j T_{i}^{AB} = 
t_i^{\mu} n^{A\nu} t_j^{\rho} \nabla_\rho \nabla_\mu n^B_{\nu}
+
\tilde{\Gamma}^m_{ij} T_m^{AB}
+
K_{il}^B K_{jk}^A h^{lk}\nn\\
&\qquad\qquad
+
T_i^{BC} T_j^{AD} \eta_{CD}
-
K_{ij}^D \Theta_D^{\;BA} 
\label{Liei3} ~,\\
&\mathcal{L}_i \Theta^{ABC} 
=
n^{A\mu} n^{C\nu} t^{\rho}_i \nabla_\rho \nabla_\mu n^B_{\nu}
-
K_{il}^{(A} T_{k}^{B C)} h^{kl} \nn
\\
&\qquad\qquad\;\;\; -T_i^{AD} \Theta^{EBC} \eta_{DE}
-
T_i^{CD} \Theta^{AB}_{\;\;\;\;\;D}
\label{Lien4} 
\end{align}

\caption{Summary of tangential Lie derivatives of the connection forms from Table~\ref{tablecf}.}\label{tabletangent}.
\end{table}

\begin{table}[h]
\centering
\begin{align}
&\mathcal{L}_A \tilde{\Gamma}_{ij}^k 
=
t^\mu_i t^{k\nu} n^{A\rho} \nabla_\rho \nabla_\mu t_{j\nu}
+
\tilde{\Gamma}_{lj}^k K_{im}^A h^{ml}
-
\tilde{\Gamma}_{ij}^m K_{ml}^A h^{kl}\nn\\
&\qquad\;\;\;\;- T^{AB}_{[i} K_{jl]}^C h^{kl} \eta_{BC}
\label{Lien1} ~, \\
&\mathcal{L}_B K_{ij}^A = 
t^\mu_i t^{\nu}_j n^{B\rho} \nabla_\rho \nabla_\mu n^A_{\nu}
+
K_{(il}^A K_{kj)}^B h^{lk}
+
T_{(i}^{AC} T_{j)}^{BD} \eta_{CD}
\label{Lien2} ~, \\
&\mathcal{L}_C T_{i}^{AB} =n^{A\mu} t^{\nu}_i n^{C\rho} \nabla_\rho \nabla_\mu n^B_{\nu}
+T_j^{A[B} K_{il}^{C]} h^{jl}+\Theta^{CA}_{\;\;\;\;\;D} T_i^{DB}+\nn\\
&\qquad\qquad\;\Theta^{AB}_{\;\;\;\;\;D} T_i^{DC}
\label{Lien3} ~, \\
&\mathcal{L}_D \Theta^{ABC} =
n^{A\mu} n^{C\nu} n^{D\rho} \nabla_\rho \nabla_\mu n^B_{\nu}
-
T_i^{(A B}T_j^{C)D} h^{ij}
+
\Theta^{DC}_{\;\;\;\;\;E} \Theta^{ABE}\nn
\\
&\qquad\qquad\;\;+\Theta^{DA}_{\;\;\;\;\;E} \Theta^{EBC}
\label{Lien4} 
\end{align}
\caption{Summary of normal Lie derivatives of the connection forms from Table~\ref{tablecf}.}\label{tablenormal}
\end{table}

Now we turn to the consequences of the normal derivatives of connections reported in Table~\ref{tablenormal}.
Notice that we can antisymmetrize~\eqref{Lien3} in $A\leftrightarrow C$ and derive the identity
\begin{align}
\mathcal{L}_{[C} T_{i}^{A]B} =&\
R_i^{\;\;BCA}
+
T_j^{[AB} K_{il}^{C]} h^{jl}
+
T_j^{[CA]} K_{il}^{B} h^{jl}\nn
\\&+
\Theta^{[CA]}_{\;\;\;\;\;D} T_i^{DB}
+
\Theta^{[AB}_{\;\;\;\;\;D} T_i^{DC]} \,.
\label{Rclie2}
\end{align}
Similarly, from~\eqref{Lien4} we get by subtracting the same equation with $A\leftrightarrow D$ 
\begin{align}
\mathcal{L}_{[D} \Theta^{A]BC} 
=&\
R^{BCAD}
-
T_i^{[A B}T_j^{CD]} h^{ij}
-
T_i^{C B}T_j^{[AD]} h^{ij}\nn\\
&+
\Theta^{[DC}_{\;\;\;\;\;E} \Theta^{A]BE}
+
\Theta^{[DA]}_{\;\;\;\;\;E} \Theta^{EBC} \,.
\label{todonormal}
\end{align}
\noindent
From the orthogonality of tangent and normal vectors we have that
\be 
t_i^\mu n^{B\rho} \nabla_\rho \nabla_\mu \left(n^A_\nu t_j^\nu\right)=0 ~,
\ee
 which implies
\begin{align}
t_i^\mu t_j^\nu n^{B\rho} \nabla_\rho \nabla_\mu n^A_\nu
=&\
-\,t_i^\mu  n^{A\nu} n^{B\rho} \nabla_\rho \nabla_\mu t_{j\nu}
+
T_k^{AB} \tilde{\Gamma}_{ij}^k
\nn\\
&+\Theta^{BA}_{\;\;\;\;\;C} K_{ij}^C
-
K_{ik}^A K_{jl}^B h^{kl}
-
T_i^{AC} T_j^{BD} \eta_{CD} \,.
\label{ttnNNn}
\end{align}
Analogously, from $n^A_\nu t_i^\mu \nabla_\mu \left(\mathcal{L}_B t_j^\nu\right)=0$ and the first Bianchi identity we get 
\begin{align}
t_i^\mu  n^{A\nu} n^{B\rho} \nabla_\rho \nabla_\mu t_{j\nu}
=&\,
t_i^\mu n^{A\nu} t_j^\rho  \nabla_\rho \nabla_\mu n_{\nu}^B
+
R^{A\;\;B}_{\;\;\;i\;\;\;j}
+
T_k^{AB} \tilde{\Gamma}_{ij}^k
-\nn\\
&\Theta^{CBA} K_{ij}^D \eta_{CD}
+
K_{ik}^B K_{jl}^A h^{kl}
+
T_i^{BC} T_j^{AD} \eta_{CD} \,.
\label{tnnNNt}
\end{align}
Finally, combining~\eqref{ttnNNn} and~\eqref{tnnNNt} we get
\begin{align}
t_i^\mu t_j^\nu n^{B\rho} \nabla_\rho \nabla_\mu n^A_\nu
=&
-
t_i^\mu n^{A\nu} t_j^\rho  \nabla_\rho \nabla_\mu n_{\nu}^B
-
R^{A\;\;B}_{\;\;\;i\;\;\;j}
+
\Theta^{[CB]A} K_{ij}^D \eta_{CD}
\nn\\
&-K_{(ik}^A K_{j)l}^B h^{kl}
-
T_{(i}^{AC} T_{j)}^{BD} \eta_{CD}  \,,
\end{align}
so that combining this with~\eqref{Liei3} we have the following identity 
\begin{align}
t_i^\mu t_j^\nu n^{B\rho} \nabla_\rho \nabla_\mu n^A_\nu
=&
-
\tilde{\nabla}_j T^{AB}_i
-
R^{A\;\;B}_{\;\;\;i\;\;\;j}
-
K_{ik}^A K_{jl}^B h^{kl}\nn\\
&-
T_{i}^{AC} T_{j}^{BD} \eta_{CD} 
-
K_{ij}^D  
\Theta^{BCA}
\eta_{CD}
\,.
\end{align}
In this way we can rewrite the normal variation of $K_{ij}^A$ as
\begin{align}
\mathcal{L}_B K_{ij}^A = &
-
\tilde{\nabla}_j T^{AB}_i
-
K_{ij}^D  
\Theta^{BCA}
\eta_{CD}
-
R^{A\;\;B}_{\;\;\;i\;\;\;j} \nn\\
&+
K_{jl}^A K_{ik}^B h^{lk}
+
T_{j}^{AC} T_{i}^{BD} \eta_{CD}
\label{LieKfull}\,.
\end{align}
The usefulness of this expression will become apparent in the following Appendix.

There are still a few invariants which will be relevant for our computations, and for which we would like to compute normal variations.
These are constructed with curvature tensors.
First, the Riemann contracted with four normal vectors:
\be
\mathcal{L}_C R_{AB}^{\;\;\;\;\;AB}
=
4 T_k^{CA}  R^{k\;\;\;\;\;B}_{\;\;BA}
+
n_A^\mu n_B^\nu n^{A\rho} n^{B\sigma} n^{C\delta} \nabla_\delta R_{\mu\nu\rho\sigma} \,.
\ee
Also,
\be
\mathcal{L}_B R_{A}^{\;\;A}
=
2 T_k^{CA}  R^{k}_{\;\;A}
+
n_A^\mu n_A^\nu n^{C\delta} \nabla_\delta R_{\mu\nu} \,.
\ee
The variation of the Ricci scalar is
\be
\mathcal{L}_B R = n^{B\rho} \nabla_\rho R  \,.
\ee
Moreover, the intrinsic Ricci scalar varies as $\mathcal{L}_B \mathcal{R}=2  K^B_{ij} \mathcal{R}^{ij}$, which upon using the contracted Gauss equation~\eqref{GClie1} becomes
\be
\mathcal{L}_B \mathcal{R}
=
2
\left[
K^B_{ij} R^{ij}
-
K^B_{ij} R^{iAj}_{\;\;\;\;\;A}
+
\tr (K^B K^A) \tr K_A
-
\tr (K^B K^A K_A )
\right]\nn
\,.
\ee

%
\section{Derivation of shape equations}\label{eom}

In this appendix we derive the Euler--Lagrange equations of motion associated with a generic functional of the form
\begin{align}\label{fnl}
S_\mathrm{eff}[\Sigma]  = \int_\Sigma d^p\sigma\ \sqrt{h}&\big[ \lambda_0 + \big(\lambda_1 \mathcal{R} + \lambda_2 R + \lambda_3 R_A^{\;\;\,A}+ \lambda_4 R_{A B}^{\;\;\;\;\;AB} \nn \\ &+ \lambda_5 \tr K_A \tr K^A+ \lambda_6 \tr K^A K_A \big) \big]\,.
\end{align}
Central to our approach is to write these equations purely in terms of geometric objects, such as $\tr K^A$, ${\cal R}$, and so on.
As usual, to find the Euler--Lagrange equations associated to a functional one needs to consider variations.
In the present case, we must consider variations of the surface $\Sigma$.
Hence, it is necessary to posses the appropriate language to discuss the geometry in vicinity of $\Sigma$, the techniques required to do so where developed in \ref{sec:app geometry}.
In principle, we must vary $\Sigma$ in all the possible directions inside $M$ however, one can show that the variations in the directions tangent to $\Sigma$  can be reabsorbed as diffeomorphisms.
Thus, we are left to consider variations in the normal direction only. 

Normal variations are implemented by the map
\be
x^\mu
\to y^\mu=
x^\mu +
\varepsilon^A(\sigma^i) n^\mu_A   \,,
\label{normaldefor}
\ee
where $\varepsilon^A:\Sigma\to \mathbb{R}$, for each $A$, is infinitesimally small and we refer to the set of points $y^\mu$ as $\Sigma'$.
Notice the similarity of the above expression with~\eqref{defining y}, which in fact corresponds to the constant $\varepsilon^A(\sigma^i) $ case; hereafter, we refer to this case as a \emph{rigid normal variation}.
With this terminology, we could say that in \ref{sec:app geometry} we learned how the geometric structures on $\Sigma$ transform under a rigid normal variation.

\begin{figure}[h]
\centering
\includegraphics[width=.77\textwidth]{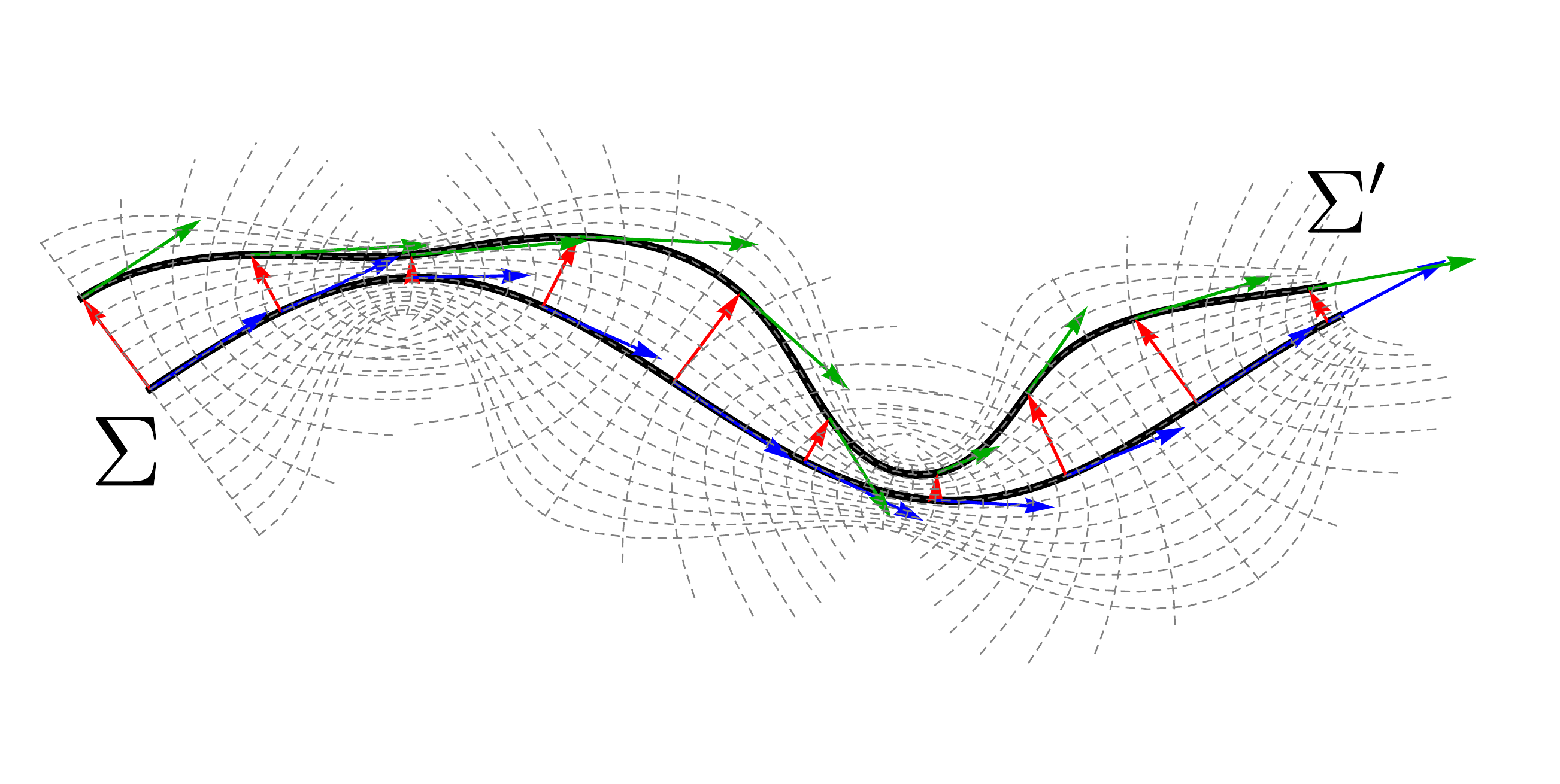}
\caption{Adapted frame.}
\label{fig1a}
\end{figure}

We wish to extend this technology to \emph{local normal variations}, \textit{i.e.}, for  $\varepsilon^A(\sigma^i) $  non-constant.
Most of the structure found in the previous appendix is preserved but there is an important subtlety to bear in mind.  Recall that the first step to perform the variations is to extend the tangent and normal bundles from $\Sigma$ to a neighborhood around it. If we extend the tangent vectors such that 
\be\label{Lie loca}
\mathcal{L}_n t_i^\mu
=0\,,
\ee
where $n^\mu=\varepsilon^A(\sigma)n_A^\mu$, then we have
\be\label{Lie loca2}
t^\nu_i \nabla_\nu n^{\mu}=  n^{\nu}\nabla_\nu t^\mu_i \,.
\ee
 Moreover, extending the normal vectors in a way which preserves orthogonality with the tangent vectors implies
\begin{align}\label{Orto}
t^\nu_in^\mu\nabla_\mu n^{A}_{\,\nu}=- n^{A}_{\,\nu}n^\mu\nabla_\mu t^\nu_i\,.
\end{align}
Combining \eqref{Lie loca2} and  \eqref{Orto} it follows that 
\be\label{shifted}
t^\nu_in^\mu\nabla_\mu n^{A}_{\,\nu}=T^{BA}_i\varepsilon_B-\partial_i\varepsilon^A \,.
\ee
Observe that this identity is shifted with respect to the rigid case, compare with the last expression in Table \ref{tablecf}.

To see the implications of the above discussion let us consider the local variation of the extrinsic curvatures. First, using Eq.\,\eqref{Lie loca} we have
\be
{\cal L}_n \, K_{ij}^A=t^\mu_it^\nu_j\,{\cal L}_n \nabla_{\mu}n_\nu^A\,.
\ee
After some manipulations, this equation can be rewritten as 
\begin{align}
{\cal L}_n \, K_{ij}^A=&\varepsilon_B\,R^{A\;\;B}_{\;\;j\;\;\;i}+t^\mu_i\nabla_\mu\left(t^\nu_jn^\lambda\nabla_\lambda\,n_\nu^A\right)-\left(t^\mu_i\nabla_\mu t^\nu_j\right)\left(n^\lambda\nabla_\lambda n_\nu^A\right) \nonumber \\
&+\varepsilon_B\left(t^\mu_j\nabla_\mu n^{B\lambda}\right)\left(t_i^\nu\nabla_\nu n^A_\lambda\right)+ n^{B\lambda}\left(t_i^\nu\nabla_\nu n^A_\lambda\right) \partial_j\varepsilon_B\,.
\end{align}
Now, we proceed to insert the identities listed in Table \ref{tablecf} into the above equation. Nevertheless, one must be careful to use the shifted identity \eqref{shifted}. This procedure yields
\bea
\mathcal{L}_n K^A_{ij}
&=&
\varepsilon_B \mathcal{L}_B K_{ij}^A
-
T_{(i}^{AB} \partial_{j)} \varepsilon_B
-
\tilde{\nabla}_i \tilde{\nabla}_j \varepsilon^A  \,,
\eea
where the $ \mathcal{L}_B K_{ij}^A$ is given by Eq.\,\eqref{LieKfull}. Following a similar procedure we can show 
\be
\mathcal{L}_n T_i^{AB}
=
\varepsilon_C \mathcal{L}_C T_i^{AB}
+
K_{ij}^{[A}  \partial^j \varepsilon^{B]} 
-
\Theta^{CAB} \partial_i \varepsilon_C   \,,
\label{Lnormalt}
\ee
where $\mathcal{L}_C T_i^{AB}$ is computed in~\eqref{Lien3}.

Now, we are set to compute the variations of the quadratic invariants appearing in the functional~\eqref{fnl}.
The variation of the square of the trace of the extrinsic curvatures is
\begin{align}
\mathcal{L}_n  \tr K_B\tr K^B
=&
- 2 \tr K_B
\Big[
\varepsilon_A \tr\left( K^B K^A\right)
+ 
\tilde{\Delta} \varepsilon^B
+
\varepsilon_A
R^{B\;\;Ai}_{\;\;i}+
2 T_i^{BA} \tilde{\nabla}^i \varepsilon_A\nn\\
&
+
\varepsilon_A\tilde{\nabla}^i T_i^{BA} 
-
\varepsilon_A h^{ij} T_i^{BC} T_j^{AD}\eta_{CD}
\Big] \,.
\end{align}
Meanwhile that of the trace of the squares reads
\begin{align}
\mathcal{L}_n  \tr \left( K_B K^B\right)
=&
-  2
\varepsilon_A
\Big[ 
\tr\left( K^B K_B K^A\right)
+K_B^{ij}  R^{B\;\;A}_{\;\;\;j\;\;\;i}
+
K_B^{ij} \tilde{\nabla}_i T_j^{BA} \\
&
-
K_B^{ij} T_i^{BC} T_j^{AD} \eta_{CD}
\Big]
-
2  K_A^{ij} \tilde{\nabla}_i \tilde{\nabla}_j \varepsilon^A
-
4 K_B^{ij} T_i^{BA}  \tilde{\nabla}_j \varepsilon_A \,.\nn
\end{align}
Finally, the variations of the contractions of the Riemann tensor are given by
\begin{align}
&\mathcal{L}_n   R_{CB}^{\;\;\;\;\;CB} 
=
4 \left(\varepsilon_A T_k^{AC} - \partial_k \varepsilon^C \right) R^{k\;\;\;\;\;B}_{\;\;BC}
+
\varepsilon_A n_C^\mu n_B^\nu n^{C\rho} n^{B\sigma} n^{A\delta} \nabla_\delta R_{\mu\nu\rho\sigma}\nn \,,
\\
&\mathcal{L}_n 
R_{B}^{\;\;B}
=
2 \left(\varepsilon_A T_k^{AC} -\partial_k \varepsilon^C \right)  R^{k}_{\;\;C}
+
\varepsilon_A n_C^\mu n^{C\nu} n^{A\delta} \nabla_\delta R_{\mu\nu} \,.
\end{align}
Bringing these things together and integrating by parts we find the Euler--Lagrange equations of motion
\be
\lambda_0\tr K^A+\sum_{n=1}^6\,\lambda_n {\cal E}^A_n=0 ~, \label{eq:eom1 B}
\ee
where
\begin{align}
& {\cal E}^A_1=\tr K^A{\cal R}-2{\cal R}^{ij}K^A_{ij}, \\
& {\cal E}^A_2=\tr K^A R+ n^{A}_{\mu}\nabla^\mu R , \\
& {\cal E}^A_3=\tr K^A R_{B}^{\;\;B}+2\tilde D_k^{\,AB}R^k_{\,B}+ n_C^\mu n^{C\nu} n^{A\delta} \nabla_\delta R_{\mu\nu},\\
 &{\cal E}^A_4=\tr K^A   R_{CB}^{\;\;\;\;\;CB}+4\tilde D_k^{\,AB}R^{kC}_{\;\;\;BC}+ n_C^\mu n_B^\nu n^{C\rho} n^{B\sigma} n^{A\delta} \nabla_\delta R_{\mu\nu\rho\sigma}, \\
& {\cal E}^A_5=  \tr K_B\left[\tr K^A\tr K^B -2\tr\left(K^B  K^A\right)-2R^{B\;\;Ai}_{\;\;i}\right]\\
&\hspace{9mm}-2\tilde D_{i\;\;C}^{\;A}\tilde D^{iCB} \tr K_B,\nn \\
& {\cal E}^A_6=-  2
\left[ \tilde D_{i\;\;B}^{\;A} \tilde D_j^{\;BC} K_C^{ij}+\tr \left( K^B K_B K^A\right)+
K_B^{ij}  R^{B\;\;A}_{\;\;\;j\;\;\;i}
\right]\\
&\hspace{9mm}+\tr K^A  \tr \left( K_B K^B\right),\nn
\end{align}
where we used the differential operator $\tilde D_i^{\;AB}$ defined in~\eqref{covariantDofK}.

As a closing remark for this section, let us show how normal Killing directions trivialize the normal bundle. Suppose that one of the normal directions to $\Sigma$ is the projection onto $T\Sigma$ of a Killing vector field, i.e. it exists an $\bar{A}$ such that
\be
\mathcal{L}_{\bar{A}} g_{\mu\nu} = \nabla_\mu n^{\bar{A}}_\nu + \nabla_\nu n^{\bar{A}}_\mu = 0 \,.
\label{Killingcondition}
\ee
This automatically implies that $\mathcal{L}_{\bar{A}} h_{ij} =0$ and thus in the direction $n^{\bar{A}\mu}$ the extrinsic curvature is zero
\be
K_{ij}^{\bar{A}} =0  \,.
\ee
Moreover \eqref{Killingcondition} implies also that $t_{i}^\mu \mathcal{L}_{\bar{A}} n^B_\mu=0$ which is equivalent to requiring
\be
T^{\bar{A}B}_i =0  \,,
\ee
for fixed normal index $\bar{A}$. The curvature equations imply also further constraint on projections of the Riemann tensor, explicitly:
\be 
R^{\bar{A}}_{\;\;i jk }
=
R^{\bar{A}B}_{\;\;\;\;\;ij}
=
R^{\bar{A}BC}_{\;\;\;\;\;\;\;\;i} =0 \,. 
\ee
Summarizing, every time we can find a normal vector field on $\Sigma$ which is also a Killing for $M$, we can de facto reduce the codimension of the problem.
For example, for time-independent space-times, whenever one is considering static embeddings, the time-like direction is always a Killing vector field and the problem can be reduced in finding extrema of \eqref{S2action} in a static foliation of $M$.

\section{Inverting $\tr K$ in maximally symmetric surfaces}\label{sec:trK}
In this Appendix, we carry out in detail the strategy outlined in~\cite{Langer1984} to invert the extrinsic curvature in maximally symmetric spaces.
In Section~\ref{sec:curves} we saw that the equations for a curve in a maximally symmetric space can be written as an equation for $\tr K$.
See, for example,~\eqref{Curve in MSS}.
Leaving the codimension arbitrary, extrema must satisfy
\be\label{trk eq}
2\tilde\Delta\tr K^A+\tr K^A \tr K_B \tr K^B-\left(\frac{\hat\lambda_0}{\lambda'_5}-\frac{2\kappa}{L^2}\right) \tr K^A=0 ~,
\ee
in a torsionless frame. Notice that, locally, such frame can always be found for curves.
Using the arclength parametrization, contracting with $\dot K_A$ and integrating, we find
\be\label{al}
\dot K_A\,\dot K^A +K_A K^A \left(\frac{\kappa}{L^2} -\frac{\hat\lambda_0}{2\lambda_5'}+\frac{1}{4} K_B K^B \right) = \mathrm{constant} \,,
\ee
where  $\dot\,=d/ds$.
After solving this equation along the lines of Section~\ref{sec:curves}, we are left with the task of inverting $\tr K^A$ to find the sought after extrema, $\Sigma$.
The first step to attain this goal is to construct a Killing vector field along an extremal curve $\Sigma$, \textit{i.e.}, we must find  
\be
w^\mu(\sigma) = w_{||}(\sigma) t^\mu + w^A_\perp(\sigma) n_A^\mu \,,
\ee 
such that
\be \label{kill}
\mathcal{L}_w h= 0 
\qquad \mathrm{and}\qquad
\mathcal{L}_w K^A = 0 \,,
\ee
for a solution of~\eqref{al}.
The conditions~\eqref{kill} are equivalent to stating that $w^\mu$ is a Killing vector field in a neighborhood of $\Sigma$.
The first condition implies
\be
\dot w_{||}+  w^A_\perp(\sigma) K_A = 0 \,,
\ee
while the latter yields
\be
  w_{||}(\sigma)\dot  K^A - \left(  w^B_\perp(\sigma) K_B K^A +  \ddot  w^A_\perp(\sigma)+ \frac{\kappa}{L^2} w^A_\perp(\sigma)
\right) 
=0
\,.
\ee
Both of these are solved by
\be
w^\mu 
=
\left(K_A K^A -\lambda\right) t^\mu 
-
2\dot K^A\, n_A^\mu \,,
\label{killingw}
\ee
where $\lambda=\hat\lambda_0/\lambda_5'$.
 The crucial point is that since we are in a maximally symmetric space, where the number of isometries is maximal, any local Killing field must originate as the restriction of a global Killing field
\be
w^\mu =\left. \omega^\mu \right|_\Sigma  \,.
\ee 
Hence, $\omega^\mu$ provides a natural extension of $w^\mu $ to the whole ambient space. 

Below, we study some aspects of the integral curves associated to $\omega^\mu$, which we refer to generically as $\gamma_\omega$.
As curves in their own right, the $\gamma_\omega$ induce a natural decomposition of the tangent space into the tangent vector and its orthogonal complement.
For $\Sigma$ this decomposition is the familiar $\{t^\mu, n^\mu_A\}$.
In turn for the $\gamma_\omega$, we have $\{\omega^\mu, m^\mu_a\}$, with $a=1,\dots, d-1$.
At the points where $\Sigma$ and $\gamma_\omega$ meet, these two bases provide alternative descriptions of the tangent space.
Just as with  $\Sigma$, every $\gamma_\omega$  has an induced metric $h_\omega=\omega^\mu\omega_\mu$ and an extrinsic curvature for each of its normal directions
\be\label{k omega}
K^a_\omega=\omega^\mu \omega^\nu \nabla_\mu m^a_\nu\,.
\ee
(The last expression must not be confused with $K^A$~\eqref{Extrinsic curvature} corresponding to $\Sigma$.)
Since $\omega^\mu$ is a Killing field we can show that 
\be
{\cal L}_{\omega}K^a_\omega=0\,.
\ee
Thus, we find that the $\gamma_\omega$ are constant mean curvature curves.

Now, we express the value of~\eqref{k omega} on $\Sigma$ in terms of the curvatures $K^A$.
For simplicity, we consider the case $d=2$ but the result can be readily generalized.
For this case, we have a single $K^A$ which we denote by $k$.
The same applies to $K^a_\omega$ which we write as $k_\omega$.
On $\Sigma$ we can decompose $\left. \omega^\mu \right|_\Sigma=w^\mu$ in the  $\{t^\mu, n^\mu_A\}$ basis.
Introducing  $w_t=w_\mu t^\mu$ and $w_n=w_\mu n^\mu$, we can write~\eqref{k omega} as
\be 
 \left.\tr\, k_\omega \right|_\Sigma
=
\frac{w_t^3  k
-
w_t^2
\dot w_n
+
w_t w_n
\left[
\dot w_t
-
\mathcal{L}_n w_n
\right]
+
w_n^2
\mathcal{L}_n w_t 
}{\left[w_t^2+w_n^2\right]^{3/2}} \,.
\ee 
Finally, using~\eqref{killingw} to find $w_t$ and $w_n$ as well as the formula~\eqref{Lien2}, we obtain
\be
 \left.\tr\, k_\omega \right|_\Sigma
=
-\frac{2k}{\sqrt{\left(k^2-\lambda\right)^2+4\dot k\,^2}}
\left(
\frac{\kappa}{L^2} 
+
\frac{4  k^3 \dot k^2}
{\left(k^2-\lambda\right)^2+4\dot k\,^2} 
\right)\,.
\label{curvatureofomegap1}
\ee
This equation leads to some interesting observations.
First, it implies that if at an intersection point between $\gamma_\omega$ and $\Sigma$ the extrinsic curvature $k$ of $\Sigma$ vanishes, then $k_\omega=0$ at that point as well.
Furthermore, we have argued that the curves $\gamma_\omega$ are constant mean curvature solutions.
Hence, $k_\omega=0$ all along the curve, and $\gamma_\omega$ is in fact a geodesic.\footnote{
Interestingly, if $\lambda=0$ then this geodesic and $\Sigma$ intersect orthogonally.}
Similarly, if $ \dot k=0$ at the intersection point, then $\gamma_\omega$ is a constant mean curvature solution with 
\be\label{inter CMC}
 \tr\, k_\omega =-\frac{2\kappa k}{L^2|k^2-\lambda|}\,.
\ee
Notice that in flat space ($\kappa=0$) these curves are also geodesics.

\subsection{Extrema in $\mathbb{H}^2$}\label{sec:wavy}

We consider extrema in the two dimensional Lobachevsky space $\mathbb{H}^2$.
We can study this space in two representations, the Poincar\'e disk and the upper half plane.
The former has coordinates $\{r,\phi\}$ while the latter has $\{z,x\}$.
Setting the radius of curvature $L=1$, the line element reads
\be
ds^2 = \frac{4}{\left(1-r^2\right)^2} \left(dr^2+r^2 d\phi^2\right)=\frac{1}{z^2}\left(dx^2+dz^2\right) \, ,
\ee
in the Poincar\'e disk and the upper half plane, respectively. Hereafter we focus on the latter. 
Now, consider a curve in $\mathbb{H}^2$ parametrized by arclength.
The tangent and normal vectors are given by
\be
t^\mu(s)=\left(\dot z,\sqrt{z^2-\dot z^2}\right)\qquad n^\mu(s)=\pm\left(\sqrt{z^2-\dot z^2},-\dot z\right)\,,
\ee
and the extrinsic curvature reads
\be\label{k uhp}
k(s)=\pm\frac{1}{z}\left(\dot z+\frac{\dot z\,^2-z \ddot z}{\sqrt{z^2-\dot z^2}} \right)\,.
\ee
In the following, we consider the reparametrization
\be \label{new param}
z(s) = e^{-f(s)} \,.
\ee

Now, we employ the Killing field technology developed in the previous section to find extrema in $\mathbb{H}^2$.
Imagine that we have found the extrinsic curvature $k(s)$ of the $\Sigma$ by solving~\eqref{trk eq} and assume that $k(s)$  has at least one zero.
Then, from the discussion following~\eqref{curvatureofomegap1}, we know that at the point where $k(s)$ vanishes $\Sigma$ must intersect an integral geodesic of the Killing field $\omega^\mu$.
Generically, a Killing vector field in $\mathbb{H}^2$ can be written as
\be \label{Kill H2}
\omega^\mu= c_1  \left(0,1\right)+c_2 \left(z,x\right)+c_3  \left(2x z,x^2-z^2\right) \,,
\ee
where the three vectors correspond to translations, dilatations, and special conformal transformations.
To any vector field of the form~\eqref{Kill H2} we can associate a unique integral geodesic 
\be
c_1 + c_2 x -c_3 ( z^2 + x^2) = 0  \,.
\ee 
Hence, at the zeroes of $k(s)$,  $\Sigma$ must intersect one of the above curves.

Without loss of generality, we choose the geodesic $x=0$ which can be mapped to the other geodesics via isometries.
In the arclength parametrization, this geodesic has $z(s)=e^{-s}$; its corresponding Killing vector is
\be 
\omega^\mu\propto\left( z,x\right)\,.
\ee
Comparing this with~\eqref{killingw}, we find the system
\begin{align}
(k^2 -\lambda) \dot z- 2\dot k \sqrt{z^2-\dot z^2} &=  {\cal A} z \,,\label{eq t}\\
(k^2 -\lambda) \sqrt{z^2-\dot z^2}+ 2 \dot k \dot z&={\cal A}x \,,
\label{geodesiceom}
\end{align}
where ${\cal A}$ is a normalization constant.
In the coordinate~\eqref{new param}, we discover that~\eqref{eq t} becomes
\be 
\dot f(s) = -\frac{{\cal A}(k^2-\lambda) \pm 2 \dot k\sqrt{(k^2-\lambda)^2+4\dot k^2-{\cal A}^2}}{(k^2-\lambda)^2+4\dot k^2} \,.
\label{fprime} 
\ee
Meanwhile, imposing $\omega^\mu \omega_\mu = w^\mu w_\mu$ implies
\be
(k^2-\lambda)^2+4\dot k^2={\cal A}^2\left(1+\frac{x^2}{z^2}\right) \,,
\label{geodesiceommodulus}
\ee
which on the geodesic $x=0$ becomes 
\be \label{Aa}
{\cal A}=\sqrt{\lambda^2+4\,\dot k^2(s_0)} \,,
\ee
where $s_0$ is the arclength value at which $\Sigma$ intersects the geodesic $x=0$.
The constant ${\cal A}$ is related to the right hand side of~\eqref{al}, which now reads
\be
4 \dot k^2 +(k^2-\lambda)^2- \frac{4 k^2 }{L^2 }= {\cal A}^2   \,.
\label{integratedeomH2}
\ee
This last expression, together with~\eqref{geodesiceommodulus} leads to a succinct relation between $x(s)$ and $z(s)$:
\be\label{xz}
\left(\frac{x}{z}\right)^2 =  \frac{4k^2}{{\cal A}^2 L^2}  \,.
\ee
Using~\eqref{integratedeomH2}, we can simplify~\eqref{fprime} to state
\be 
\dot f(s) = -\frac{{\cal A}(k^2-\lambda) \pm \frac{4 k \dot k}{L}}{{\cal A}^2+ \frac{4 k^2}{L^2}} \,.
\label{fprimetointegrate}
\ee

\begin{figure}[h!]
\centering
\includegraphics[scale=0.3]{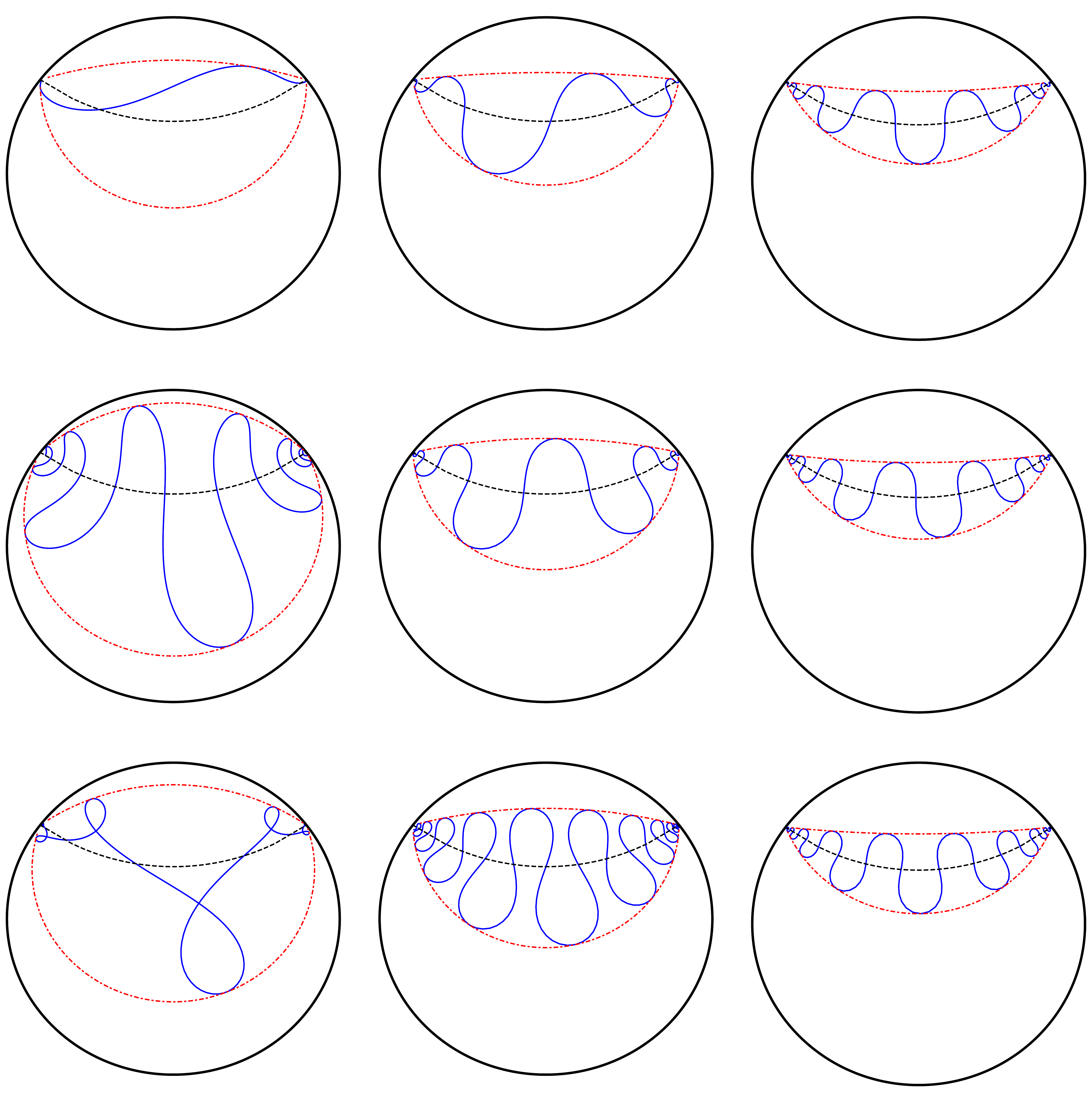}
\caption{Examples of wavelike solutions Eq.\,\eqref{wavy explicit} in the Poincar\'e disk. Each wavelike solution is accompanied by its generating geodesic (black) and its two bounding CMCs \eqref{bounding} (red). The parameters used in this plot are: rows, top to bottom $\lambda=-1.5$, $\lambda=0$ and $\lambda=1.5$; columns, left to right $A=0.5$, $A=20$ and $A=200$. }
\label{fig:wavy}
\end{figure}

Recapitulating, to find extrema in $\mathbb{H}_2$,  first we must find the extrinsic curvature by solving~\eqref{trk eq}.
The relevant solutions are discussed in Section~\ref{sec:curves}.
Once we have found the extrinsic curvature $k(s)$, we integrate~\eqref{fprimetointegrate} to find $f(s)$, from which $z(s)$ can be determined.
Finally, we find $x(s)$ using~\eqref{xz} and we have thus found the extrema $\Sigma$.
Following this line of thought we will find the curve corresponding to the wavelike solution~\eqref{op1}
\be
k(s)=\sqrt{2+\lambda+ C}\;\mathrm{cn}\left(\sqrt{\frac{C}{2}}s,\frac{C+2+\lambda}{2C}\right)\,,
\label{wavyH2}
\ee
where we defined $C=\sqrt{A+(2+\lambda)^2}$ and assumed $A>0$.
The next step is to find the zeroes of~\eqref{wavyH2}, which occur at
\be
s_l=\sqrt{\frac{2}{C}}(2l+1) K\Big(\frac{C+2+\lambda}{2C}\Big)\,,\qquad \, l\in \mathbb{N} \,,
\ee
where $K(m)$ is an elliptic integral of the first kind~\eqref{complete ell}.
The constant~\eqref{Aa} can be determined from any of these zeroes:
\be \label{calA}
{\cal A} = \sqrt{C^2-4(\lambda+1)} \,.
\ee
Moreover, at the critical points of $k(s)$ from~\eqref{inter CMC} we find the curvature of the constant mean curvature solutions that intersect $\Sigma$
\be \label{bounding}
\tr\, k_\omega =
\frac{2\sqrt{2+\lambda+C}}{2+C}
\,.
\ee
In the upper half-plane model these curves correspond to straight lines originating at $(z,x)=(0,0)$ with slope
\be
\pm \frac{{\cal A}}{2\sqrt{2+\lambda+C}} \,. 
\ee
Finally, plugging~\eqref{wavyH2} into~\eqref{fprimetointegrate} and integrating, we find
\be\label{wavy explicit}
z(s)
=
\frac{C}{2+\lambda}
\frac{
\exp\left[\sqrt{C^2-4(\lambda+1)}\left(\frac{s}{4}-\frac{2(C-2)}{4\sqrt{2C}(C+2)}\Pi\left[n,\varphi(s); m
\right]\right)\right]
}{\sqrt{(C+2)^2-4(C+2+\lambda)\mathrm{sn}\,^2\left(\sqrt{\frac{C}{2}}s,\frac{C+2+\lambda}{2C}\right)}} \,,
\ee
where $\Pi$ is the incomplete elliptic integral of the third kind \eqref{Pi} and we have
\be
n=\frac{4(C+2+\lambda)}{(C+2)^2}\;,\qquad m=\frac{C+2+\lambda}{2C}\,,
\ee
and
\be
\varphi(s)=\mathrm{amp}\,
\left( 
\sqrt{\frac{C}{2}}s
,
\frac{C+2+\lambda}{2C}
\right)\,.
\ee
 Finally, $x(s)$ can be obtained using~\eqref{xz}.

\section{Jacobi elliptic functions}\label{sec:jac}

For the reader's convenience, we recall the definitions of special functions used throughout the paper.
The elliptic functions may be constructed from the incomplete elliptic integral of the first kind:
\be
z = F(\varphi,m) = \int_0^\varphi \frac{d\theta}{\sqrt{1-m\sin^2\theta}} ~.
\ee
Here, the elliptic modulus $m$ satisfies $0<m<1$.
The amplitude is
\be
\varphi = F^{-1}(z,m) = \am(z,m) ~.
\label{eq:amp}
\ee
We then have
\be
\sn(z,m) = \sin\varphi ~, \;\;\;
\cn(z,m) = \cos\varphi ~, \;\;\;
\dn(z,m) = \sqrt{1-m\sin^2\varphi} ~.
\label{eq:jacobi}
\ee
These are doubly periodic generalizations of the trigonometric functions:
\bea
&& \sn(z+2\ell K+2niK',m) = (-1)^\ell\, \sn(z,m) ~, \nn \\
&& \cn(z+2\ell K+2niK',m) = (-1)^{\ell+n}\, \cn(z,m) ~, \\
&& \dn(z+2\ell K+2niK',m) = (-1)^n\, \dn(z,m) ~, \nn
\eea
where $\ell$ and $n$ are integers and $K$ and $K'$ are defined from the complete elliptic integral of the first kind:
\be\label{complete ell}
K = K(m) = \int_0^{\frac{\pi}{2}} \frac{d\theta}{\sqrt{1-m\sin^2\theta}} = \frac{\pi}{2} {}_2F_1(\frac12,\frac12;1;m^2) ~, 
\ee
and
\be
K' = K(m') = K(\sqrt{1-m}) ~.
\ee
From the definitions~\eref{eq:jacobi}, the Jacobi elliptic functions satisfy the identities
\be
\sn^2(z,m)+\cn^2(z,m) = 1 ~, \qquad
m\,\sn^2(z,m)+\dn^2(z,m) = 1 ~.
\label{eq:sqid}
\ee
Special values include
\bea
\sn(z,0) = \sin z ~, & \quad \sn(z,1) = \tanh z ~, & \quad \cn(z,0) = \cos z ~, \nn \\
\cn(z,1) = \mathrm{sech}\, z ~, & \quad \dn(z,0) = 1 ~, & \quad \dn(z,1) = \mathrm{sech}\, z ~.
\eea
Using the Glaisher notation, we express reciprocals and quotients as
\bea
&& \mathrm{ns}(z,m) = \frac{1}{\sn(z,m)} ~, \quad \mathrm{nc}(z,m) = \frac{1}{\cn(z,m)} ~, \quad \mathrm{nd}(z,m) = \frac{1}{\dn(z,m)} ~, \nn \\
&& \mathrm{sc}(z,m) = \frac{\sn(z,m)}{\cn(z,m)} ~, \quad \mathrm{sd}(z,m) = \frac{\sn(z,m)}{\dn(z,m)} ~, \quad \mathrm{cd}(z,m) = \frac{\cn(z,m)}{\dn(z,m)} ~,\nn \\
&& \mathrm{cs}(z,m) = \frac{\cn(z,m)}{\sn(z,m)} ~, \quad \mathrm{ds}(z,m) = \frac{\dn(z,m)}{\sn(z,m)} ~, \quad \mathrm{dc}(z,m) = \frac{\dn(z,m)}{\cn(z,m)} ~. \nn
\eea
Finally, we introduce the incomplete elliptic integral of the second kind:
\be\label{incom}
E(\varphi,m) = \int_0^\varphi d\theta\ \sqrt{1-m\sin^2\theta}\,.
\ee
The complete elliptic integral of the second kind is $E(m) = E(\frac\pi2,m)$.
The incomplete elliptic integral of the third kind is
\begin{align}\label{Pi}
\Pi(n;\varphi,m)& = \int_0^\varphi \frac{d\theta}{(1-n\sin^2\theta)\sqrt{1-m^2\sin^2\theta}}\nn \\
&= \int_0^{\sin\varphi} \frac{dt}{(1-n t^2)\sqrt{(1-t^2)(1-m^2t^2)}} ~.
\end{align}
The complete elliptic integral of the third kind is $\Pi(n,m) = \Pi(n;\frac\pi2,m)$.

The general solution to the differential equation~\eref{Curve in MSS arc2} is~\eref{eq:jns}:
\be
u(s) = \alpha \left[ 1 - \frac{\alpha-\beta}{\alpha} \mathrm{ns}^2(\frac12\sqrt{\alpha-\beta}\, s+\delta, \,\frac{\alpha-\gamma}{\alpha-\beta}) \right] ~,
\label{eq:jns}
\ee
with $\delta$ a free parameter.
Setting $\delta = i K(\sqrt{1-m})$, we can rewrite $\mathrm{ns}(z-\delta,m) = m\, \sn(z,m)$.
This enables us to massage~\eref{eq:jns} to read
\be
u(s) = \alpha \left[1-\frac{\alpha-\gamma}{\alpha} \sn^2(\frac12\sqrt{\alpha-\beta}\, s,\,\frac{\alpha-\gamma}{\alpha-\beta}) \right] ~.
\ee
We therefore recover~\eref{eq:u-of-s}.

\newpage 

\bibliography{biblio}{}

\end{document}